\documentclass[english,notitlepage,longbibliography,showpacs,preprintnumbers,amsmath,amssymb,aps,prx,nofootinbib,12pt,superscriptaddress]{revtex4-1}
\usepackage[T1]{fontenc}
\usepackage[latin9]{inputenc}
\usepackage[paper=a4paper,top=1in,bottom=1in,right=1in,left=1in]{geometry} 
\usepackage{color}
\usepackage{xcolor}
\usepackage{verbatim}
\usepackage{units}
\usepackage{amsmath}
\usepackage{amsthm}
\usepackage{amssymb}
\PassOptionsToPackage{normalem}{ulem}
\usepackage{ulem}
\usepackage{lipsum}

\usepackage{scrlayer-scrpage} 
\usepackage{textcomp} 

\usepackage[pdftex]{graphicx}

\makeatletter
\usepackage{microtype}
\usepackage{xspace}
\usepackage{amsfonts}
\usepackage{palatino}
\usepackage[]{algorithm2e}
\usepackage{MnSymbol}
\usepackage{bbm}
\usepackage{url}
\usepackage{ragged2e}
\edef\UrlBreaks{\do\-\UrlBreaks}
\usepackage{accents} 
\usepackage{microtype}

\usepackage{color}

\let\OLDthebibliography\thebibliography 
\renewcommand\thebibliography[1]{
  \OLDthebibliography{#1}
  \setlength{\parskip}{0pt}
  \setlength{\itemsep}{0pt plus 3.3ex}
}

\newlength{\bibitemsep}
\newlength{\bibparskip}
\let\oldthebibliography\thebibliography
\renewcommand\thebibliography[1]{%
  \oldthebibliography{#1}%
  \setlength{\parskip}{\bibitemsep}%
  \setlength{\itemsep}{\bibparskip}%
}

\newcommand*{\mathcolor}{}
\def\mathcolor#1#{\mathcoloraux{#1}}
\newcommand*{\mathcoloraux}[3]{%
  \protect\leavevmode
  \begingroup
    \color#1{#2}#3%
  \endgroup
}
\usepackage{color}
\definecolor{dred}{rgb}{.8,0.2,.2}
\definecolor{ddred}{rgb}{.8,0.5,.5}
\definecolor{dblue}{rgb}{.2,0.2,.8}
 
\makeatletter
\let\@msm@th@eqref\eqref
\renewcommand{\eqref}[1]{%
  \begingroup
  \leavevmode
  \color{violet}%
  \hypersetup{linkbordercolor=[named]{violet}}%
  \@msm@th@eqref{#1}%
  \endgroup
}
\makeatother

\def\eqref#1{\textcolor{black}{(\ref{#1})}}

\usepackage{enumitem} 
\setlist[itemize]{topsep=0pt}

\usepackage[colorlinks,urlcolor=blue,bookmarksopen=true]{hyperref}
\usepackage{bookmark}

\newcommand{\secspace}{\vspace{2mm}}

\newcommand{\summarysec}{\secspace\emph{\textbf{Summary}}}
\newcommand{\prossec}{\secspace\emph{\textbf{Pros}}}
\newcommand{\conssec}{\secspace\emph{\textbf{Cons}}}
\newcommand{\costsec}{\secspace\emph{\textbf{Cost}}}
\newcommand{\examplesec}{\secspace\emph{\textbf{Example}}}
\newcommand{\refsec}{\secspace\emph{\textbf{Bibliography}}}
\newcommand{\altnamesec}{\secspace\emph{\textbf{Alternate Names}}}
\newcommand{\altformsec}{\secspace\emph{\textbf{Alternate Forms}}}

\makeatother

\usepackage{babel}
\begin{document}

\vspace{-40mm}
\newgeometry{top=0.85in,left=0.75in,right=0.75in}

\begin{titlepage}
\title{\vspace{-2.0cm}Quadratization in Discrete Optimization and Quantum Mechanics}






\author{\vspace{1.5mm}Nike Dattani}
\email{nik.dattani@gmail.com}

\begin{abstract}
\vspace{4mm} 
An open source Book on quadratizations for classical computing, quantum annealing, and universal adiabatic quantum computing. 

\vspace{-5mm}  

\noindent 
\end{abstract}

\maketitle

\title{\tableofcontents{}}

\noindent\rule{\textwidth}{0.4pt}
\vspace{-1.25mm} 

\indent When optimizing discrete functions, it is often easier when the function is quadratic than if it is of higher degree. But notice that the cubic and quadratic functions:

\vspace{-4mm}

\begin{align}
b_1b_2+b_2b_3+b_3b_4 -4b_1b_2b_3 & \hspace{5mm} \textrm{(cubic)},\label{eq:intro_cubic}\\
b_1b_2 + b_2b_3 + b_3b_4 + 4b_1 -4b_1b_2 -4b_1b_3 & \hspace{5mm}\textrm{(quadratic)},\label{eq:intro_quadratic}
\end{align}

\vspace{1mm}

\noindent where each $b_i$ can either be 0 or 1, both never go below the value of -2, and all minima occur at $(b_1,b_2,b_3,b_4) = (1,1,1,0)$.  Therefore if we are interested in the ground state of a discrete function of degree $k$, we may optimize either function and get exactly the same result. { \textbf{Part \ref{partDiagonal}}} gives more than 40 different ways to do this, almost all of them published in the last 5 years. 


\vspace{0.5mm}
\noindent\rule{\textwidth}{0.4pt}
\\
\vspace{-2mm}
\\
\indent The binary variables $b_i$ can be either of the eigenvalues of the matrix $b$ below, which is related to the Pauli $z$ matrix by $z=2b-\openone$. The Pauli matrices $x,y,z,\openone$ are listed below:
\vspace{0mm}
\begin{equation}
b\equiv\begin{pmatrix}1 & 0\\
0 & 0
\end{pmatrix},\,z\equiv\begin{pmatrix}1 & 0\\
                0 & -1
                \end{pmatrix},\, x\equiv\begin{pmatrix}0 & 1\\
1 & 0\end{pmatrix},\,y\equiv\begin{pmatrix}0 & -\textrm{i}\\
                      \textrm{i} & 0
                     \end{pmatrix},\,\openone\equiv\begin{pmatrix}1 & 0\\ 
                                      0 & 1
                     \end{pmatrix}.  \label{eq:pauli}
\end{equation}  

\noindent Any Hermitian $2\times 2$ matrix can be written as a linear combination of the Pauli matrices, so we can therefore describe the Hamiltonian of any number of spin-$\nicefrac{1}{2}$ particles by a function of Pauli matrices acting on each particle, for instance:

\vspace{-5mm}

\begin{align}
x_1y_2z_3y_4 + y_1x_2z_3y_4 + x_1x_2y_3 & \hspace{5mm} \textrm{(cubic)},\label{eq:intro_quantum_cubic}\\
x_1y_4 + x_2y_4 + x_3 & \hspace{5mm} \textrm{(quadratic)},\label{eq:intro_quantum_quadratic}
\end{align}

\noindent where the coefficients tell us about the strengths of couplings between these particles. The Schr\"{o}dinger equation tells us that the eigenvalues of the Hamiltonian are the allowed energy levels and their eigenvectors (wavefunctions) are the corresponding physical states. More generally these do not have to be spins but can be any type of qubits, and we can encode the solution to \textit{any} problem in the ground state of a Hamiltonian, then solve the problem by finding the lowest energy state of the physical system (this is called adiabatic quantum computing). Eqs. \eqref{eq:intro_quantum_quadratic} and \eqref{eq:intro_quantum_cubic} have exactly the same energy spectra, so Eq. \eqref{eq:intro_quantum_quadratic} is an example of a type of quadratization.

Two-body physical interactions occur more naturally than many-body interactions so \textbf{Parts \ref{partTransverseIsing}-\ref{partGeneral}} give more than 30 different ways to quadratize general Hamiltonians (some of these methods may use $d\times d$ matrices instead of only the $2\times2$ matrices in Eq. \eqref{eq:pauli}, meaning that we can have types of qudits that are not qubits). All of these methods were published during the last 15 years.

 The optimization problems of Eqs. \eqref{eq:intro_cubic}-\eqref{eq:intro_quadratic} are specific cases of the type in Eqs. \eqref{eq:intro_quantum_cubic}-\eqref{eq:intro_quantum_quadratic}, but with only $b$ matrices.

\noindent\rule{\textwidth}{0.4pt}
\end{titlepage}
\addtocounter{page}{1}

\restoregeometry

\newpage

\part{{\normalsize{\underline{Diagonal Hamiltonians (pseudo-Boolean functions)}}}\label{partDiagonal}} 

\section{Methods that introduce ZERO auxiliary variables}

\subsection{Deduction Reduction (Deduc-reduc; Tanburn, Okada, Dattani, 2015)} \label{subsec:deduc_reduc}

\vspace{-5mm}

\summarysec

We look for \emph{deductions} (e.g. $b_{1}b_{2}=0$) that must hold true at the global minimum.
These can be found by \emph{a priori} knowledge of the given problem, or by enumerating solutions of a small subset of the variables.
We can then substitute high-order terms using the low-order terms of the deduction, and add on a penalty term to preserve the ground states \cite{Tanburn2015c}.

\costsec
\begin{itemize}
\item $0$ auxiliary variables needed.
\item For a particular value of $m$, we have $\binom{n}{m}$ different $m$-variable subsets of the $n$ variable problem, and $\binom{n}{m} 2^m$ evaluations of the objective function to find all possible $m$-variable deductions, whereas $2^n$ evaluations is enough to solve the entire problem. We therefore choose $m \lll n$.
\end{itemize}

\prossec
\begin{itemize}
\item No auxiliary variables needed.
\end{itemize}

\conssec
\begin{itemize}
\item When deductions cannot be determined naturally (as in the Ramsey number determination problem, see Example \ref{subsec:Example_Ramsey_deduc_reduc}), deductions need to be found by `brute force', which scales exponentially with respect to $m$.
For highly connected systems (systems with a large number of non-zero coefficients), the value of $m$ required to find even one deduction can be prohibitively large.
\end{itemize}

\examplesec

Consider the objective function:
\vspace{-2mm}
\begin{equation}
H_{4{\rm -local}}=b_{1}b_{2}(4+b_{3}+b_{3}b_{4})+b_{1}(b_{3}-3)+b_{2}(1-2b_{3}-b_{4})+F(b_{3},b_{4},b_{5},\ldots,b_{N})
\end{equation}
where $F$ is any quadratic polynomial in $b_{i}$ for $i\ge 3$.
Since
\vspace{-2mm}
\begin{equation}
H_{4{\rm -local}}\left(1, 1, b_3, b_4, ...\right) > H_{4{\rm -local}}\left(0, 0, b_3, b_4, ...\right), H_{4{\rm -local}}\left(0, 1, b_3, b_4, ...\right), H_{4{\rm -local}}\left(1, 0, b_3, b_4, ...\right),
\end{equation}
\uline{\textbf{\textit{it must be the case that $b_{1}b_{2}=0$}}}.
Specifically, for the 4 assignments of $(b_{3},b_{4})$, we see that $b_{1}b_{2}=0$ at every minimum of $H_{4{\rm -local}}-F$.

Using deduc-reduc we have:
\vspace{-2mm}
\begin{equation}
H_{2{\rm -local}}=6b_{1}b_{2}+b_{1}(b_{3}-3)+b_{2}(1-2b_{3}-b_{4})+F(b_{3},b_{4},b_{5},\ldots,b_{N}),
\end{equation}
which has the same global minima as $H_{4-\textrm{local}}$ but one fewer quartic and one fewer cubic term.
The coefficient of $b_1 b_2$ was chosen as $6$ because $6 \ge \max \left( 4+b_{3}+b_{3}b_{4} \right)$.

\refsec
\begin{itemize}
\item Original paper, with more implementation details, and application to integer factorization: \cite{Tanburn2015c}.
\end{itemize}

\newpage

\subsection{ELC Reduction (Ishikawa, 2014)}

\summarysec

An Excludable Local Configuration (ELC) is a partial assignment of variables that make it impossible to achieve the minimum.
We can therefore add a term that corresponds to the energy of this ELC without changing the solution to the minimization problem.
In practice we can eliminate all monomials with a variable in which a variable is set to 0, and reduce any variable set to 1.
Given a general objective function we can try to find ELCs by enumerating solutions of a small subset of variables in the problem \cite{Ishikawa2014}. 

\costsec
\begin{itemize}
\item $0$ auxiliary variables needed.
\item For a particular value of $m$, we have $\binom{n}{m}$ different $m$-variable subsets of the $n$ variable problem, and $\binom{n}{m} 2^m$ evaluations of the objective function to find all possible $m$-variable deductions, whereas $2^n$ evaluations is enough to solve the entire problem. We therefore choose $m \lll n$.
\item Approximate methods exist which have been shown to be much faster and give good approximations to the global minimum \cite{Ishikawa2014}.
\end{itemize}

\prossec
\begin{itemize}
\item No auxiliary variables needed.
\end{itemize}

\conssec
\begin{itemize}
\item No known way to find ELCs except by `brute force', which scales exponentially with respect to $m$. 
\item ELCs do not always exist.
\end{itemize}

\examplesec

Consider the objective function: 
\begin{equation}
H_{{\rm 3-local}}=b_{1}b_{2}+b_{2}b_{3}+b_{3}b_{4}-4b_{1}b_{2}b_{3}. \label{eq:ELCexample}
\end{equation}
If $b_{1}b_{2}b_{3}=0$, no assignment of our variables will we be able to reach a lower energy than if $b_{1}b_{2}b_{3}=1$.
Hence this gives us \textit{twelve} ELCs, and one example is $(b_{1},b_{2},b_{3})=(1,0,0)$ which we can use to form the polynomial:
\begin{align}
H_{2\rm{-local}}&=H_{{\rm 3-local}}+4b_{1}(1-b_{2})(1-b_{3})\\
&=b_{1}b_{2}+b_{2}b_{3}+b_{3}b_{4}+4b_{1}-4b_{1}b_{2}-4b_{1}b_{3}. \label{eq:ELCreduced}
\end{align}
In both cases Eqs. \eqref{eq:ELCexample} and \eqref{eq:ELCreduced}, the only global minima occur when $b_1b_2b_3 = 1$.

\refsec
\begin{itemize}
\item Original paper and application to computerized image denoising: \cite{Ishikawa2014}.
\end{itemize}

\newpage
\subsection{Groebner Bases}

\summarysec

Given a set of polynomials, a Groebner basis is another set of polynomials that have exactly the same zeros.
The advantage of a Groebner basis is it has nicer algebraic properties than the original equations, in particular they tend to have smaller degree polynomials.
The algorithms for calculating Groebner bases are generalizations of Euclid's algorithm for the polynomial greatest common divisor.

Work has been done in the field of 'Boolean Groebner bases', but while the variables are Boolean the coefficients of the functions are in $\mathbb{F}_{2}$ rather than $\mathbb{Q}$.

\costsec
\begin{itemize}
\item $0$ auxiliary variables needed.
\item $\mathcal{O}\left( 2^{2^{n}} \right)$ in general, $\mathcal{O}(d^{n^{2}})$ if the zeros of the equations form a set of discrete points, where $d$ is the degree of the polynomial and $n$ is the number of variables \cite{Bardet2002}.
\end{itemize}

\prossec
\begin{itemize}
\item No auxiliary variables needed.
\item General method, which can be used for other rings, fields or types of variables.
\end{itemize}

\conssec
\begin{itemize}
\item Best algorithms for finding Groebner bases scale double exponentially in $n$.
\item Only works for objective functions whose minimization corresponds to solving systems of discrete equations, as the method only preserves roots, not minima.
\end{itemize}

\examplesec

Consider the following pair of equations:

\begin{equation}
b_1 b_2 b_3 b_4 + b_1 b_3 + b_2 b_4 - b_3 = b_1 + b_1 b_2 + b_3 - 2 = 0.
\end{equation}

\noindent Feeding these to Mathematica's ${\tt GroebnerBasis}$ function, along with the binarizing $b_1(b_1-1)=\ldots=b_4(b_4-1)=0$ constraints, gives a Groebner basis:

\begin{equation}
\left\{ b_4 b_3 - b_4, b_2 + b_3 - 1, b_1 - 1 \right\}.
\end{equation}

From this we can immediately read off the solutions $b_1=1$, $b_2=1-b_3$ and reduce the problem to $b_3b_4-b_4=0$. Solving this gives a final solution set of: $(b_1, b_2, b_3, b_4) = (1, 0, 1, 0), (1, 0, 1, 1), (1, 1, 0, 0)$, which should be the same as the original 4-local problem.

\refsec
\begin{itemize}
\item Reduction and embedding of factorizations of all bi-primes less than $200,000$: \cite{Dridi2016}.
\end{itemize}

\newpage

\subsection{Application of ELM (Dattani, 2018)}

\vspace{-3mm}

\summarysec

We use the formula from \cite{Ali2008} for representing any function of three binary variables:
\vspace{-4mm}

{\footnotesize
\begin{align}
f(b_1,b_2,b_3) &= \left( f(1,1,1) + f(1,0,0) - f(1,1,0) - f(1,0,1) - f(0,1,1) - f(0,0,0) \right. + \\
               & \left.   f(0,0,1) + f(0,1,0)  \right)b_1b_2b_3 + \left(f(0,1,1) + f(0,0,0) - f(0,0,1) - f(0,1,0)  \right)b_2b_3 ~+ \\
               & \left( f(1,0,1) + f(0,0,0) - f(0,0,1) - f(1,0,0) \right)b_1b_3 + \\
               & \left(f(1,1,0) + f(0,0,0) - f(1,0,0) - f(0,1,0)  \right)b_1b_2 ~+ \left( f(0,1,0) - f(0,0,0) \right)b_2  \\
               &  + \left(f(1,0,0) - f(0,0,0)\right)b_1 + \left(  f(0,0,1) -\left. f(0,0,0)  \right)b_3 + f(0,0,0).\right.
\end{align}
}

\vspace{-4mm}

If the cubic term is zero, then the function becomes quadratic. We can use ELM (Energy Landscape Manipulation) to change the energy landscape \textit{without} changing the ground state \cite{Tanburn2015d}. In this case, we apply ELM in order to make the cubic term zero.

\costsec
\begin{itemize}

\item No auxiliary variables.
\item May require many evaluations of the cubic function, varying coefficients in order to find the right ELM coefficients.
\item For a particular value of $m$, we have $\binom{n}{m}$ different $m$-variable subsets of the $n$ variable problem, and $\binom{n}{m} 2^m$ evaluations of the objective function to find all possible $m$-variable deductions, whereas $2^n$ evaluations is enough to solve the entire problem. We therefore choose $m \lll n$.
\end{itemize}

\prossec
\begin{itemize}
\item Can be generalized to arbitrary $k$-local functions, but the ELM constraints may become harder to achieve.
\item Can quadratize an entire cubic function (or a cubic part of a more general function) with no auxiliary qubits.
\item Can reproduce the full spectrum.
\end{itemize}

\conssec
\begin{itemize}
\item May not always be possible.
\item May require a local search to find appropriate deductions.
\end{itemize}

\examplesec

In order to reduce the number of constraints required for the cubic term to be zero, we will assume that Deduc-Reduc told us that $(1-b_1)(1-b_2) + (1-b_2)(1-b_3) + (1-b_1)(1-b_3) = 0$ when the overall function is minimized, which means the ground state only occurs when at least two variables are 1. This does not allow us to assign the linear terms, but assigns all quadratic terms to have $b_ib_j=1$. This also suggests that $b_1b_2b_3$ can also be reduced to a linear term, but we do not know whether it is $b_1$, $b_2$, or $b_3$. So we have the following constraint on the cubic term, after setting all $f(b_1,b_2,b_3)$ to zero if there is not at least two 1's:

\begin{align}
f(1,1,1) - f(1,1,0) - f(1,0,1) - f(0,1,1) = 0.
\end{align}


\refsec
\begin{itemize}
\item The method was first presented in the first arXiv version of this book \cite{Dattani2019}.
\end{itemize}

\newpage

\subsection{Split Reduction (Okada, Tanburn, Dattani, 2015)}

\summarysec

It is possible to reduce a lot of the problem by conditioning on the most connected variables.
We call each of these operations a \emph{split}.

\costsec

Usually slightly sub-exponential in the number of splits, as the number of problems to solve at most doubles with every split, but often does not double (since entire cases can get eliminated by some splits, as in the example below).

\prossec
\begin{itemize}
\item This method can be applied to any problem and can be very effective on problems with a few very connected variables.
\end{itemize}

\conssec
\begin{itemize}
\item Multiple runs of the optimization procedure need to be made, and the number of runs can often grow  almost exponentially with respect to the number of splits.
\end{itemize}

\examplesec

Consider the simple objective function 
\begin{equation}
H=1+b_{1}b_{2}b_{5}+b_{1}b_{6}b_{7}b_{8}+b_{3}b_{4}b_{8}-b_{1}b_{3}b_{4}.
\end{equation}
In order to quadratize $H$, we first have to choose a variable over which to split.
In this case $b_{1}$ is the obvious choice since it is present in the most terms and contributes to the quartic term.

We then obtain two different problems:
\begin{eqnarray}
H_{0} & = & 1+b_{3}b_{4}b_{8}\\
H_{1} & = & 1+b_{2}b_{5}+b_{6}b_{7}b_{8}+b_{3}b_{4}b_{8}-b_{3}b_{4}.
\end{eqnarray}

\noindent At this point, we could split $H_{0}$ again and solve it entirely, or use a variable we saved in the previous split to quadratize our only problem.

To solve $H_{1}$, we can split again on $b_{8}$, resulting in two quadratic problems:
\begin{eqnarray}
H_{1,0} & = & 1+b_{2}b_{5}-b_{3}b_{4}\\
H_{1,1} & = & 1+b_{2}b_{5}+b_{6}b_{7}.
\end{eqnarray}
Now both of these problems are quadratic.
Hence we have reduced our original, hard problem into 3 easy problems, requiring only 2 extra (much easier) runs of our minimization algorithm, and without needing any auxiliary variables.

\vspace{5mm}

Note that the number of quadratic problems to solve is 3, which is smaller than 2$^2$ which would be the "exponential" cost if the number of problems were (hypothetically) to double with each split. This is a good example of the typical \textit{\textbf{sub}}-exponential scaling of split-reduc.

\refsec
\begin{itemize}
\item Original paper and application to Ramsey number determination: \cite{Okada2015}.
\end{itemize}

\newpage
\section{Methods that introduce auxiliary variables to quadratize a SINGLE negative term (Negative Term Reductions,  NTR)}

\subsection{NTR-KZFD (Kolmogorov \& Zabih, 2004; Freedman\& Drineas, 2005)} \label{subsec:Negative-Monomial-Reduction}

\summarysec

For a negative term $-b_{1}b_{2}...b_{k}$, introduce a single auxiliary variable $b_a$ and make the substitution:
\begin{equation}
-b_{1}b_{2} \ldots b_{k} \rightarrow (k-1)b_a - \sum_ib_ib_a.
\end{equation}

\costsec
\begin{itemize}
\item 1 auxiliary variable for each $k$-local term.
\end{itemize}

\prossec
\begin{itemize}
\item All resulting quadratic terms are submodular (have negative coefficients).
\item Can reduce arbitrary order terms with only 1 auxiliary.
\item Reproduces the full spectrum.
\end{itemize}

\conssec
\begin{itemize}
\item Only works for negative terms.
\end{itemize}

\examplesec

\begin{align}
H_{{\rm 6-local}} & =-2b_{1}b_{2}b_{3}b_{4}b_{5}b_{6} + b_5b_6,
\end{align}
has a unique minimum energy of -1 when all $b_i=1$.

\begin{equation}
H_{{\rm 2-local}}=2\left( 5b_a-b_{1}b_a-b_{2}b_a-b_{3}b_a-b_{4}b_a-b_{5}b_a-b_{6}b_a \right) + b_5b_6
\end{equation}
has the same unique minimum energy, and it occurs at the same place (all $b_i=1$), with $b_a=1$.

\altformsec

\begin{align}
-b_{1}b_{2} \ldots b_{k} &= \min_{b_a} \left( (k-1-\sum_i b_{i})b_a \right)\\
&\rightarrow \left( (k-1-\sum_i b_{i})b_a \right).
\end{align}

\altnamesec 
\begin{itemize}
\item "Standard quadratization" of negative monomials \cite{Anthony2017}.
\item $s_k(b,b_a)$ \cite{Anthony2017}
\end{itemize}

\refsec
\begin{itemize}
\item 2004: Kolmogorov and Zabih presented this for cubic terms \cite{Kolmogorov2004}.
\item 2005: Generalized to arbitrary order by Freedman and Drineas \cite{Freedman2005}.
\item Discussion: \cite{Ishikawa2011}, \cite{Anthony2016}.
\end{itemize}

\newpage

\subsection{NTR-ABCG (Anthony, Boros, Crama,  Gruber, 2014)} \label{subsec:Negative-Monomial-Reduction-2}

\summarysec



For a negative term $-b_{1}b_{2}...b_{k}$, introduce a single auxiliary variable $b_a$ and make the substitution:
\begin{equation}
-b_{1}b_{2} \ldots b_{k} \rightarrow \sum_{i}^{k-1}b_{i}-\sum_{i}^{k-1}b_{i}b_{k}-\sum_{i}^kb_{i}b_{a}+(k-1)b_{k}b_{a}.
\end{equation}

\costsec
\begin{itemize}
\item 1 auxiliary variable for each $k$-local term.
\item 1 non-submodular term for each $k$-local termi (and it is quadratic).
\end{itemize}

\prossec
\begin{itemize}
\item Can reduce arbitrary order terms with only 1 auxiliary.
\item Reproduces the full spectrum.
\end{itemize}

\conssec
\begin{itemize}
\item Only works for negative terms.
\item Turns a symmetric term into a non-symmetric term (but only $b_{k}$ is asymmetric).
\end{itemize}

\examplesec

\begin{align}
H_{{\rm 6-local}} & =-2b_{1}b_{2}b_{3}b_{4}b_{5}b_{6} + b_5b_6,
\end{align}
has a unique minimum energy of -1 when all $b_i=1$.

$H_{2-\textrm{local}}$ has the same unique minimum energy, and it occurs at the same place (all $b_i=1$), with $b_a=1$.

\altformsec
\begin{align}
-b_{1}b_{2}\ldots b_{k}&\rightarrow(k-1)b_{k}b_{a}-\sum_{i}b_{i}(b_{a}+b_{k}-1)\\
                       & = (k-2)b_kb_a - \sum_i^{k-1}b_i\left( b_a + b_k -1 \right)
\end{align}

\altnamesec 
\begin{itemize}
\item "Extended standard quadratization" of negative monomials (see Eqs. 25-26 of \cite{Anthony2017}).
\item $s_k(b,b_a)^+$ \cite{Anthony2017}.
\end{itemize}

\refsec
\begin{itemize}
\item 2014, First presentation: \cite{Anthony2014,Anthony2016}. 
\item Further discussion: \cite{Anthony2017}.
\end{itemize}

\newpage

\subsection{NTR-ABCG-2 (Anthony, Boros, Crama,  Gruber, 2016)} \label{subsec:Negative-Monomial-Reduction-3}

\summarysec

For a negative term $-b_{1}b_{2}...b_{k}$, introduce a single auxiliary variable $b_a$ and make the substitution:
\begin{align} \label{eqn:NTR-ABCG-2}
-b_{1}b_{2}\ldots b_{k}	&\rightarrow\left(2k-1\right)b_{a}-2\sum_{i}b_{i}b_{a}
\end{align}

\costsec
\begin{itemize}
\item 1 auxiliary variable for each $k$-local term.
\item 1 non-submodular term for each $k$-local term (and it is linear).
\end{itemize}

\prossec
\begin{itemize}
\item Can reduce arbitrary order terms with only 1 auxiliary.
\item Reproduces the full spectrum.
\item The non-submodular term is linear as opposed to NTR-ABCG-1 whose non-submodular term is quadratic.
\item Symmetric with respect to all non-auxiliary variables.
\end{itemize}

\conssec
\begin{itemize}
\item Only works for negative terms.
\item Turns a symmetric term into a non-symmetric term (but only $b_{k}$ is asymmetric).
\item Coefficients of quadratic terms are twice the size of their size in NTR-KZFD or NTR-ABCG-1, and roughly twice the size for the linear term.
\end{itemize}

\examplesec

\begin{align}
H_{{\rm 6-local}} & =-2b_{1}b_{2}b_{3}b_{4}b_{5}b_{6} + b_5b_6,
\end{align}
has a unique minimum energy of -1 when all $b_i=1$.

$H_{2-\textrm{local}}$ has the same unique minimum energy, and it occurs at the same place (all $b_i=1$), with $b_a=1$.

\altformsec 
\begin{align}
  - b_1 b_2 \dots b_k = 2 b_{a} \left(k - \frac12 - \sum_{i=1}^k b_i \right)
\end{align}

\eqref{eqn:NTR-ABCG-2} can be generalized as follows:

\begin{align} \label{eqn:NTR-ABCG-2-gen}
-b_{1}b_{2}\ldots b_{k}	&\rightarrow\left(Ck-1\right)b_{a}-C\sum_{i}b_{i}b_{a}
\end{align}
where $C \geq 1$ is a constant. NTR-KZFD is a particular case of \eqref{eqn:NTR-ABCG-2-gen} where $C = 1$.


\refsec
\begin{itemize}
\item Discussion: \cite{Anthony2016}.
\end{itemize}

\newpage

\subsection{NTR-GBP (``Asymmetric cubic reduction'', Gallagher, Batra, Parikh, 2011)}


\summarysec
\begin{align}
-b_{1}b_{2}b_{3}&\rightarrow  b_a \left( -b_1 + b_2 + b_3  \right) -b_1b_2 - b_1b_3 + b_1 \\
                &\rightarrow  b_a \left( -b_2 + b_1 + b_3  \right) -b_1b_2 - b_2b_3 + b_2 \\
                &\rightarrow  b_a \left( -b_3 + b_1 + b_2  \right) -b_2b_3 - b_1b_3 + b_3 \label{eqn:NTR-GBP,3}
\end{align}

\costsec
\begin{itemize}
\item 1 auxiliary variable per negative cubic term.
\end{itemize}

\prossec
\begin{itemize}
\item Asymmetric which allows more flexibility in cancelling with other quadratics.
\end{itemize}

\conssec
\begin{itemize}
\item Only works for negative cubic monomials.
\end{itemize}

\examplesec
\begin{eqnarray}
- b_{1}b_{2}b_{3} + b_1 b_3 - b_2 &= \min_{b_a} \left( b_a-b_{1}b_a-b_{3}b_a+b_{2}b_a+2b_{1}b_{3} \right) - b_2
\end{eqnarray}


\altformsec
\begin{align}
-b_{1}b_{2}b_{3}&= \min_{b_a} \left( b_a  -b_1 + b_2 + b_3   -b_1b_2 - b_1b_3 + b_1 \right)
\end{align}

\begin{itemize}
\item By starting with \eqref{eqn:NTR-GBP,3}, and flipping $b_a$ (i.e. setting $b_a \rightarrow 1- \bar{b}_a$ and relabelling $b_a \rightarrow \bar{b}_a$ since $b_a$ does not appear anywhere else in the function being quadratized), we see that NTR-GBP can actually be derived from NTR-ABCG with $k=3$).
\end{itemize}

\refsec
\begin{itemize}
\item First introduced in: \cite{Gallagher2011}.
\end{itemize}

\newpage

\subsection{NTR-RBL (Rocchetto, Benjamin, Li, 2016)}

\summarysec

Using a ternary variable $t_q \in {-1,0,1}$ we have:

\begin{equation}
-z_1z_2z_3 \rightarrow \left(1 + 4t_a + z_1 + z_2 + z_3\right)^2 -1.
\end{equation}

\costsec

\begin{itemize}
\item 1 auxiliary ternary variable 
\end{itemize}

\prossec
\begin{itemize}
\item One of the only methods designed specifically for $z$ variables.
\item Symmetric with respect to all variables.
\end{itemize}

\conssec
\begin{itemize}
\item The auxiliary variable required is ternary (a qutrit).
\item Requires all possile quadratic terms and they are all non-submodular.
\item Only reproduces the ground state manifold.
\end{itemize}

\examplesec
\begin{eqnarray}
-b_{1}b_{2}b_{3}&= \min_{b_a} b_a  -b_1 + b_2 + b_3   -b_1b_2 - b_1b_3 + b_1 
\end{eqnarray}

\altformsec
\begin{eqnarray}
-z_1z_2z_3z_4\rightarrow 16\,t_a^2+4\,t_a\,\sum_{i=1}^4 z_i+2\,\sum_{i=1}^4\sum_{j>i}^4z_i\,z_j+4
\end{eqnarray}

\refsec
\begin{itemize}
\item Original paper where they introduce this gadget for the \textit{end points} in the LHZ lattice: \cite{Rocchetto2016}.
\end{itemize}

\newpage

\subsection{NTR-LHZ (Lechner, Hauke, Zoller, 2015)}

\summarysec

Extra binary or ternary variable is added to ensure that the energy of the even parity sector is zero and the energy of the odd sector is higher: 


\costsec
\begin{itemize}
\item 1 auxiliary ternary variable (see appendix for transformation to a binary variable)
\end{itemize}

\prossec
\begin{itemize}
\item One of the only methods designed specifically for $z$ variables.
\end{itemize}

\conssec
\begin{itemize}
\item Only reproduces the ground state manifold, not higher excited states.
\item Requires all possile quadratic terms and they are all non-submodular.
\item Only reproduces the ground state manifold.
\end{itemize}

\examplesec
\begin{eqnarray}
-z_1z_2z_3z_4=-16\,b_1b_2b_3b_4+8\,(b_1b_2b_3+b_1b_2b_4+b_1b_3b_4+b_2b_3b_4)-\nonumber \\
4\,(b_1b_2+b_1b_3+b_1b_4+b_2b_3+b_2b_4+b_3b_4)+2\,(b_1+b_2+b_3+b_4)-1 \nonumber \\
\rightarrow 16\,t_a^2+8\,t_a\sum_{i=1}^4 b_i+8\,\sum_{i=1}^4\sum_{j>i}^4b_i\,b_j+16
\end{eqnarray}

\altformsec
\begin{eqnarray}
-z_1z_2z_3z_4\rightarrow 16\,t_a^2+4\,t_a\,\sum_{i=1}^4 z_i+2\,\sum_{i=1}^4\sum_{j>i}^4z_i\,z_j+4
\end{eqnarray}

\refsec
\begin{itemize}
\item Original paper: \cite{Lechner2015} (Eq. 4).
\item Also given in Eq. 10 of \cite{Rocchetto2016} for $i+2<j$.
\end{itemize}

\newpage

\section{Methods that introduce auxiliary variables to quadratize a SINGLE positive term (Positive Term Reductions, PTR)}

\subsection{PTR-BG (Boros and Gruber, 2014)}

\summarysec

By considering the negated literals $\bar{b}_{i}=1-b_{i}$, we recursively apply NTR-KZFD to $b_{1}b_{2}\ldots b_{k}=-\bar{b}_{1}b_{2}\ldots b_{k}+b_{2}b_{3}\ldots b_{k}$.
The final identity is:
\begin{equation}
b_{1}b_{2}\ldots b_{k}\rightarrow \left(\sum_{i=1}^{k-2}b_{a_{i}}(k-i-1+b_{i}-\sum_{j=i+1}^{k}b_{j})\right)+b_{k-1}b_{k}
\end{equation}

\costsec
\begin{itemize}
\item $k-2$ auxiliary variables for each $k$-local term.
\end{itemize}

\prossec
\begin{itemize}
\item Works for positive monomials.
\end{itemize}

\conssec
\begin{itemize}
\item $k-1$ non-submodular quadratic terms.
\end{itemize}

\examplesec
\begin{eqnarray}
b_{1}b_{2}b_{3}b_{4} & \rightarrow {b_{a_{1}}(2+b_{1}-b_{2}-b_{3}-b_{4})+b_{a_{2}}(1+b_{2}-b_{3}-b_{4})}+b_{3}b_{4}
\end{eqnarray}

\refsec
\begin{itemize}
\item Summary: \cite{Boros2014}.
\end{itemize}

\newpage

\subsection{PTR-Ishikawa (Ishikawa, 2011)}

\summarysec

This method re-writes a positive monomial using symmetric polynomials, so all possible quadratic terms are produced and they are all non-submodular: 
\begin{equation}
b_{1}...b_{k} \rightarrow  \left( \sum_{i=1}^{n_{k}}b_{a_{i}}\left(c_{i,d}\left(-\sum_{j=1}^{k}b_{j}+2i\right)-1\right)+\sum_{i<j}b_{i}b_{j} \right)
\end{equation}
where $n_{k}=\left\lfloor \frac{k-1}{2}\right\rfloor $ and $c_{i,k}=\begin{cases}
1, & i=n_{d}\text{ and }k\text{ is odd,}\\
2, & \text{else.}
\end{cases}$

\costsec
\begin{itemize}
\item $\left\lfloor \frac{k-1}{2}\right\rfloor $ auxiliary variables for each $k$-order term
\item $\mathcal{O}(kt)$ for a $k$-local objective function with $t$ terms.
\end{itemize}

\prossec
\begin{itemize}
\item Works for positive monomials.
\item About half as many auxiliary variables for each $k$-order term as the previous method.
\item Reproduces the full spectrum.
\end{itemize}

\conssec
\begin{itemize}
\item $\mathcal{O}(k^{2})$ quadratic terms are created, which may make chimerization more costly.
\item $\frac{k(k-1)}{2}$ non-submodular terms.
\item Worse than the previous method for quartics, with respect to submodularity.
\end{itemize}

\examplesec
\begin{eqnarray}
b_{1}b_{2}b_{3}b_{4} \rightarrow (3-2b_{1}-2b_{2}-2b_{3}-2b_{4})b_a+b_{1}b_{2}+b_{1}b_{3}+b_{1}b_{4}+b_{2}b_{3}+b_{2}b_{4}+b_{3}b_{4}
\end{eqnarray}

\altformsec

For even $k$, and equivalent expression is given in \cite{Boros2018QuadratizationsOS}:

\begin{align}
b_{1}b_{2}\ldots b_{k}&\rightarrow\sum_{i}b_{i}+\sum_{ij}b_{i}b_{j}+\sum_{2i}b_{a_{2i}}\left(4i-2-\sum_{j}b_{j}\right)\\
                      &\rightarrow\sum_{i}b_{i}+2\sum_{2i}b_{a_{2i}}\left(2i-1\right)+\sum_{ij}b_{i}b_{j}-\sum_{2i,j}b_{j}b_{a_{2i}}
\end{align}

\altnamesec 
\begin{itemize}
\item "Ishikawa's Symmetric Reduction" \cite{Gallagher2011}.
\item "Ishikawa Reduction" 
\item "Ishikawa" 
\end{itemize}

\refsec
\begin{itemize}
\item Original paper and application to image denoising: \cite{Ishikawa2011}.
\item Equivalent way of writing it for even $k$, shown in \cite{Boros2018QuadratizationsOS}.
\end{itemize}
\newpage


\subsection{PTR-BCR-1 (Boros, Crama, and Rodr\'{i}guez-Heck, 2018)}

\summarysec

This is very similar to the alternative form of Ishikawa Reduction, but works for odd values of $k$, and is different from Ishikawa Reduction: 

\begin{eqnarray}
\begin{gathered}
b_{1}b_{2}\ldots b_{k} \rightarrow \sum_{i}b_{i}+\sum_{2i-1}\left(4i-3\right)b_{a_{2i-1}}+\sum_{ij}b_{i}b_{j}-\sum_{2i-1,j}b_{j}b_{a_{2i-1}}
\end{gathered}
\end{eqnarray}

\costsec
\begin{itemize}
\item Same number of auxiliaries as Ishikawa Reduction.
\end{itemize}

\prossec
\begin{itemize}
\item Same as for Ishikawa Reduction.
\end{itemize}

\conssec
\begin{itemize}
\item Same as for Ishikawa Reduction.
\item Only works for odd $k$, but for even $k$ we have an analogous method which is equivalent to Ishikawa Reduction.
\end{itemize}

\altformsec
\begin{align}
b_{1}b_{2}\ldots b_{k}&\rightarrow\sum_{i}b_{i}+\sum_{ij}b_{i}b_{j}+\sum_{2i-1}b_{a_{2i-1}}\left(4i-3-\sum_{j}b_{j}\right)
\end{align}

\refsec
\begin{itemize}
\item Original paper: \cite{Boros2018QuadratizationsOS}.
\end{itemize}

\newpage

\subsection{PTR-BCR-2 (Boros, Crama, and Rodr\'{i}guez-Heck, 2018)}

\summarysec

Let $\lceil \frac{k}{4} \rceil \le m \le \lceil \frac{k}{2} \rceil$, 

\begin{align}
\begin{gathered}
b_{1}b_{2}\cdots b_{k}	\rightarrow\alpha^{b}\sum_{i}b_{i}+\alpha^{b_{a,1}}\sum_{i}b_{a_{i}}+\alpha^{b_{a,2}}b_{a_{m}}+\alpha^{bb}\sum_{ij}b_{i}b_{j}+\alpha^{bb_{a,1}}\sum_{i}\sum_{j}^{m-1}b_{i}b_{a_{j}}+\\
	\alpha^{bb_{a,2}}\sum_{i}b_{i}b_{a_{m}}+\alpha^{b_{a,1}b_{a,1}}\sum_{ij}^{m-1}b_{a_{i}}b_{a_{j}}+\alpha^{b_{a,1}b_{a,2}}\sum_{i}^{m-1}b_{a_{i}}b_{a_{m}},
\end{gathered}
\end{align}

\noindent where:

\begin{align}
\begin{pmatrix}\alpha^{b} & \alpha^{bb_{a,1}}\\
\alpha^{b_{a,1}} & \alpha^{bb_{a,2}}\\
\alpha^{b_{a,2}} & \alpha^{b_{a,1}b_{a,1}}\\
\alpha^{bb} & \alpha^{b_{a,1}b_{a,2}}
\end{pmatrix}&=\begin{pmatrix}-\nicefrac{1}{2} & -1\\
1 & -2\\
\frac{1}{2}(n-m+n^{2}-2mn+m^{2}) & -(n-m)\\
\nicefrac{1}{2} & 4(n-m)
\end{pmatrix}.
\end{align}


\costsec

$\lceil \frac{k}{4} \rceil$ auxiliary qubits per positive monomial.

\prossec
\begin{itemize}
\item Smallest number of auxiliary coefficients that scales linearly with $k$.
\item Smaller coefficients than the logarithmic reduction.
\end{itemize}

\conssec
\begin{itemize}
\item Introduces many non-submodular terms.
\end{itemize}

\examplesec

We quadratize a quartic term with only 1 auxiliary (half as many as in PTR-Ishikawa):
\begin{eqnarray}
b_1 b_2 b_3 b_4 \rightarrow  \frac{1}{2}
\left( b_1 + b_2 + b_3 + b_4 - 2b_{a_1} \right)
\left( b_1 + b_2 + b_3 + b_4 - 2b_{a_1} - 1 \right)
\end{eqnarray}

\refsec
\begin{itemize}
\item Original appears in: Theorem 7 of \cite{Boros2018QuadratizationsOS}, and Theorem 10 of \cite{Boros2018boundsPaper}.
\end{itemize}

\newpage

\subsection{PTR-BCR-3 (Boros, Crama, and Rodr\'{i}guez-Heck, 2018)} 

\summarysec


Pick $m$ such that $k < 2^{m+1}$,

\begin{align}
b_{1}b_{2}\ldots b_{k}	&\rightarrow \alpha+\alpha^{b}\sum_{i}b_{i}+\alpha^{b_{a_{i}}}\sum_{i}2^{i-1}b_{a_{i}}+\alpha^{bb}\sum_{ij}b_{i}b_{j}+\alpha^{bb_{a}}\sum_{ij}b_{i}b_{a_{j}}+\alpha^{b_{a_{i}}b_{a_{j}}}b_{a_{i}}b_{a_{j}},
\end{align}

\noindent where,

\begin{align}
\begin{pmatrix}\alpha & \alpha^{bb}\\
\alpha^{b} & \alpha^{bb_{a}}\\
\alpha^{b_{a}} & \alpha^{b_{a}b_{a}}
\end{pmatrix}=\begin{pmatrix}\left(2^{m}-k\right)^{2} & 1\\
2\left(2^{m}-k\right) & 2^{j-1}\\
-2\left(2^{m}-k\right) & 2^{i+j-2}
\end{pmatrix}.
\end{align}

\costsec

$\lceil \log k \rceil  $ auxiliary qubits per positive monomial.

\prossec
\begin{itemize}
\item Logarithmic number of auxiliary variables.
\end{itemize}

\conssec
\begin{itemize}
\item Introduces all terms non-submodular except for the term linear in auxiliaries.
\end{itemize}

\examplesec



\begin{eqnarray}
b_1 b_2 b_3 b_4 \rightarrow  \frac{1}{2}
\left( 4 + b_1 + b_2 + b_3 + b_4 - b_{a_1} - 2b_{a_2} \right)
\left( 3 + b_1 + b_2 + b_3 + b_4 - b_{a_1} - 2b_{a_2} \right)
\end{eqnarray}

\altformsec

\begin{align}
b_{1}b_{2}\ldots b_{k}	\rightarrow\left(2^{m}-k+\sum_{i}b_{i}-\sum_{i}2^{i-1}b_{a_{i}}\right)^{2}
\end{align}


\refsec
\begin{itemize}
\item Original paper (Theorem 4, special case of Theorem 1): \cite{Boros2018QuadratizationsOS}.
\end{itemize}

\newpage

\subsection{PTR-BCR-4 (Boros, Crama, and Rodr\'{i}guez-Heck, 2018)} 

\summarysec

Pick $m$ such that $k \le 2^{m+1}$,

\begin{align}
b_1 \ldots b_k &\rightarrow   \frac{1}{2} \left( 2^{m+1}-k + \sum_ib_i - \sum^m_{i} 2^i b_{a_i} \right)
\left( 2^{m+1}-k + \sum_i b_i - \sum^m_{i} 2^i b_{a_i} - 1 \right).
\end{align}

\costsec

$\lceil \log k \rceil  - 1$ auxiliary qubits per positive monomial.

\prossec
\begin{itemize}
\item Logarithmic number of auxiliary variables.
\end{itemize}

\conssec
\begin{itemize}
\item Introduces many non-submodular terms.
\end{itemize}

\examplesec


\begin{eqnarray}
b_1 b_2 b_3 b_4 \rightarrow  \frac{1}{2}
\left( b_1 + b_2 + b_3 + b_4 - 2b_a \right)
\left( b_1 + b_2 + b_3 + b_4 - 2b_a - 1 \right)
\end{eqnarray}

\altformsec

Let $X = \sum b_i$ and $N = 2^{m+1} - k$,

\begin{align}
b_1 \ldots b_k &= \min_{b'_1, \ldots b'_n} \frac{1}{2} \left( N + X - \sum^n_{i=1} 2^i b'_i \right)
\left( N + X - \sum^n_{i=1} 2^i b'_i - 1 \right).
\end{align}

\refsec
\begin{itemize}
\item Original paper (Theorem 5): \cite{Boros2018QuadratizationsOS}, Also in (Theorem 9): \cite{Boros2018boundsPaper}.
\end{itemize}

\newpage



\subsection{PTR-BCR-5 (Boros, Crama, and Rodr\'{i}guez-Heck, 2018)} 

\summarysec

Pick $m$ such that $k \le 2^{m+1}$,

\begin{align}
b_1 \ldots b_k &\rightarrow   \frac{1}{2} \left( 2^{m+1}-k + \sum_ib_i - \sum^m_{i} 2^i b_{a_i} \right)
\left( 2^{m+1}-k + \sum_i b_i - \sum^m_{i} 2^i b_{a_i} - 1 \right).
\end{align}

\costsec

$\lceil \log n \rceil  - 1$ auxiliary qubits per positive monomial.

\prossec
\begin{itemize}
\item Logarithmic number of auxiliary variables.
\item "As mentioned in Section 1, Theorem 9 provides a significant improvement
over the best previously known quadratizations for the Positive monomial,
and the upper bound on the number of auxiliary variables precisely matches
the lower bound presented in Section 3."
\end{itemize}

\conssec
\begin{itemize}
\item Introduces many non-submodular terms.
\end{itemize}

\examplesec


\begin{eqnarray}
b_1 b_2 b_3 b_4 \rightarrow  \frac{1}{2}
\left( b_1 + b_2 + b_3 + b_4 - 2b_a \right)
\left( b_1 + b_2 + b_3 + b_4 - 2b_a - 1 \right)
\end{eqnarray}

\altformsec

Let $X = \sum b_i$ and $N = 2^{m+1} - k$,

\begin{align}
b_1 \ldots b_k &= \min_{b'_1, \ldots b'_n} \frac{1}{2} \left( N + X - \sum^n_{i=1} 2^i b'_i \right)
\left( N + X - \sum^n_{i=1} 2^i b'_i - 1 \right).
\end{align}

\refsec
\begin{itemize}
\item Original paper (Theorem 9): \cite{Boros2018boundsPaper}.
\end{itemize}

\newpage

\subsection{PTR-BCR-5 (Boros, Crama, and Rodr\'{i}guez-Heck, 2018)} 

\summarysec

Pick $m$ such that $k \le 2^{m+1}$,

\begin{align}
b_1 \ldots b_k &\rightarrow   \frac{1}{2} \left( 2^{m+1}-k + \sum_ib_i - \sum^m_{i} 2^i b_{a_i} \right)
\left( 2^{m+1}-k + \sum_i b_i - \sum^m_{i} 2^i b_{a_i} - 1 \right).
\end{align}

\costsec

$\lceil \log n \rceil  $ auxiliary qubits per positive monomial.

\prossec
\begin{itemize}
\item Logarithmic number of auxiliary variables.
\item "As mentioned in Section 1, Theorem 9 provides a significant improvement
over the best previously known quadratizations for the Positive monomial,
and the upper bound on the number of auxiliary variables precisely matches
the lower bound presented in Section 3."
\end{itemize}

\conssec
\begin{itemize}
\item Introduces many non-submodular terms.
\end{itemize}

\examplesec


\begin{eqnarray}
b_1 b_2 b_3 b_4 \rightarrow  \frac{1}{2}
\left( b_1 + b_2 + b_3 + b_4 - 2b_a \right)
\left( b_1 + b_2 + b_3 + b_4 - 2b_a - 1 \right)
\end{eqnarray}

\altformsec

Let $X = \sum b_i$ and $N = 2^{m+1} - k$,

\begin{align}
b_1 \ldots b_k &= \min_{b'_1, \ldots b'_n} \frac{1}{2} \left( N + X - \sum^n_{i=1} 2^i b'_i \right)
\left( N + X - \sum^n_{i=1} 2^i b'_i - 1 \right).
\end{align}

\refsec
\begin{itemize}
\item Original paper (Remark 5): \cite{Boros2018boundsPaper}.
\end{itemize}

\newpage

\subsection{CCG-based (Yip, Xu, Koenig and Kumar, 2019)}
\summarysec

The CCG-based quadratization algorithm is an iterative algorithm. The CCG (Constraint composite graph) is a combinatorial structure associated with an optimization problem posed as the weighted constraint satisfaction problem. 

\costsec
\begin{itemize}
\item Each positive monomial of degree $i$ generates 2 auxiliary variables when it is reduced to the sum of a quadratic polynomial and a monomial of degree $i-1$, which can then be combined with existing monomials of degree $i-1$ if they are composed of the same variables. This combination of monomials can take place in each iteration, until the whole pseudo-Boolean function (PBF) becomes quadratic.
\item Same as Ishikawa's method, use 1 auxiliary variable for each negative monomial.
\item  For a $k$-local objective function, use a factor of $k$ less of auxiliary variables asymptotically compared to Ishikawa's method.
\end{itemize}

\prossec
\begin{itemize}
\item Due to the recombinations
of terms during its iterative reduction process, the resulting number of auxiliary variables is less than Ishikawa's method especially for PBFs with many positive monomials and many terms (the difficult case).
\item The higher the degree of the PBF is,
the more advantageous the CCG-based quadratization method is. It
works particularly well for problems in real life such as planning problem that requires a high-degree PBF formulation.
\item Due to the nature of recombinations
of terms during each iteration, the number of quadratic terms in the finalized quadratic PBFs is less than Ishikawa's method.
\end{itemize}

\conssec
\begin{itemize}
\item Can lead to more auxiliary variables for sparse PBFs (the number of monomials is much less than the maximum number the PBFs can have) and PBFs with low degree.

\end{itemize}

\examplesec

Each degree-\(d\) monomial in the PBF \(f(\vec x)\) in one reduction step is substituted as:
\begin{align}
  \begin{split}\label{eq:ccgqdr}
ax_1\ldots x_d =& \min_{x_a, x_L} \left[ax_a + Lx_L + J\sum_{i=1}^{d-1} (1-x_i)(1-x_a)  \right. \\
+ &\left.    J(1-x_d)(1-x_L) +J (1-x_L)(1-x_a) \vphantom{\sum_{xxx}^{xxx}}\right]\\
- &L(1-x_d) - a + ax_1\ldots x_{d-1},
  \end{split}\\
  -ax_1\ldots x_d =&  \min_{x_{a'}} \left[ax_{a'} + J\sum_{i=1}^d (1-x_i)(1-x_{a'}) \right] - a \label{eq:ccgqdrnegative}
\end{align}
where $J \geq L > a > 0$. $x_a$ and $x_L$ are the two auxiliary variables introduced for positive monomial and $x_{a'}$ is the auxiliary variable introduced for negative monomial.

\refsec
\begin{itemize}
\item 2019, original paper: \cite{yxkk19}.
\item 2019, (Theorem 1): \cite{yxkk19}, is inspired by 2008, (Theorem 3): \cite{vchoi08}.
\end{itemize}

\newpage

\subsection{PTR-KZ (Kolmogorov \& Zabih, 2004)}

\summarysec

This method can be used to re-write positive or negative cubic terms in terms of 6 quadratic terms.
The identity is given by:
\begin{align}
b_{1}b_{2}b_{3} & \rightarrow 1  - \left( b_a + b_1 + b_2 + b_3 \right) + b_a \left( b_1 + b_2 + b_3 \right) + b_1 b_2 + b_1 b_3 + b_2 b_3
\end{align}

\costsec
\begin{itemize}
\item 1 auxiliary variable per positive or negative cubic term.
\end{itemize}

\prossec
\begin{itemize}
\item Works on positive or negative monomials.
\item Reproduces the full spectrum.
\end{itemize}

\conssec
\begin{itemize}
\item Introduces all 6 possible non-submodular quadratic terms.
\end{itemize}

\altnamesec
\begin{itemize}
\item "Reduction by Minimum Selection" \cite{Gallagher2011,Ishikawa2011}.
\end{itemize}

\refsec
\begin{itemize}
\item Original paper: \cite{Kolmogorov2004}.
\end{itemize}

\newpage

\subsection{PTR-KZ (in terms of $z$)}

\summarysec

The formula is almost the same as in the version of PTR-KZ on the previous page (which is written in terms of $b$), but with a factor of 2, a change of sign for the linear terms, and a slight change in the constant term. This formula can be obtained directly from the PTR-KZ quadratization formula in terms of $b$, by starting with $8b_1b_2b_3$ on the left-side, making the substitution $b_i \rightarrow (1+z_i)/2$, and removing all terms that appear on both sides of the equation. The result is:

\begin{align}
\pm z_1z_2z_3 &\rightarrow 3 \pm \left( z_1 +  z_2 + z_3 + z_a \right) + 2z_a \left(z_1 + z_2 +z_3 \right) +  z_1z_2 + z_1z_3 + z_2z_3.
\end{align}

\costsec
\begin{itemize}
\item 1 auxiliary variable per positive or negative cubic term.
\end{itemize}

\prossec
\begin{itemize}
\item Works on positive or negative monomials.
\item Reproduces the full spectrum.
\end{itemize}

\conssec
\begin{itemize}
\item Introduces all 6 possible non-submodular quadratic terms.
\end{itemize}

\refsec

\begin{itemize}
\item 2004: published by Kolmogorov and Zabih in terms of $b$ variables: \cite{Kolmogorov2004}.
\item 31 March 2016: published by Chancellor, Zohren, and Warburton in terms of $z$, and without the constant term: \cite{Chancellor2016b}.
\item 8 April 2016: published independently by Leib, Zoller, and Lechner in terms of $z$, and without the constant term \cite{Leib2016a,Leib2016}.
\end{itemize}

\newpage

\subsection{PTR-GBP (``Asymmetric reduction'', Gallagher, Batra, Parikh, 2011)}

\summarysec

Similar to other methods of reducing one term, this method can reduce a positive cubic monomial into quadratic terms using only one auxiliary variable, while introducing fewer non-submodular terms than the symmetric version.

The identity is given by:
\begin{align}
b_{1}b_{2}b_{3}\rightarrow   b_a-b_{2}b_a-b_{3}b_a+b_{1}b_a+b_{2}b_{3} \\
               \rightarrow   b_a-b_{1}b_a-b_{3}b_a+b_{2}b_a+b_{1}b_{3} \\
               \rightarrow   b_a-b_{1}b_a-b_{2}b_a+b_{3}b_a+b_{1}b_{2} .\\
\end{align}

\costsec

1 auxiliary variable per positive cubic term.

\prossec
\begin{itemize}
\item Works on positive monomials.
\item Fewer non-submodular terms than Ishikawa Reduction.
\end{itemize}

\conssec
\begin{itemize}
\item Only been shown to work for cubics.
\end{itemize}

\examplesec
\begin{eqnarray}
b_{1}b_{2}b_{3} + b_1 b_3 - b_2 \rightarrow  \left( b_a-b_{1}b_a-b_{3}b_a+b_{2}b_a+2b_{1}b_{3} \right) - b_2
\end{eqnarray}

\refsec
\begin{itemize}
\item Original paper and application to computer vision: \cite{Gallagher2011}.
\end{itemize}

\newpage

\subsection{PTR-RBL-(3$\rightarrow$2) (Rocchetto, Benjamin, Li, 2016)}

\summarysec

Using a ternary variable $t_q \in {-1,0,1}$ we have:

\begin{equation}
z_1z_2z_3 \rightarrow (1 + 4t_a + z_1 + z_2 + z_3)^2 -1.
\end{equation}

\costsec

\begin{itemize}
\item 1 auxiliary ternary variable 
\end{itemize}

\prossec
\begin{itemize}
\item One of the only methods designed specifically for $z$ variables.
\item Symmetric with respect to all variables.
\end{itemize}

\conssec
\begin{itemize}
\item The auxiliary variable required is ternary (a qutrit).
\item Requires all possile quadratic terms and they are all non-submodular.
\item Only reproduces the ground state manifold.
\end{itemize}

\examplesec
\begin{eqnarray}
-b_{1}b_{2}b_{3}&= \min_{b_a} b_a  -b_1 + b_2 + b_3   -b_1b_2 - b_1b_3 + b_1 
\end{eqnarray}

\altformsec
\begin{eqnarray}
-z_1z_2z_3z_4\rightarrow 16\,t_a^2+4\,t_a\,\sum_{i=1}^4 z_i+2\,\sum_{i=1}^4\sum_{j>i}^4z_i\,z_j+4
\end{eqnarray}

\refsec
\begin{itemize}
\item Original paper where they introduce this gadget for the \textit{end points} in the LHZ lattice: \cite{Rocchetto2016}.
\end{itemize}

\newpage

\subsection{PTR-RBL-(4$\rightarrow$2) (Rocchetto, Benjamin, Li, 2016)}

\summarysec

Extra binary or ternary variable is added to ensure that the energy of the even parity sector is zero and the energy of the odd sector is higher: 

\begin{equation}
z_1z_2z_3z_4\rightarrow 16\,t_a^2+4\,t_a\,\sum_{i=1}^4 z_i+2\,\sum_{i=1}^4\sum_{j>i}^4z_i\,z_j+4.
\end{equation}

\costsec
\begin{itemize}
\item 1 auxiliary ternary variable 
\end{itemize}

\prossec
\begin{itemize}
\item One of the only methods designed specifically for $z$ variables.
\end{itemize}

\conssec
\begin{itemize}
\item Only reproduces the ground state manifold, not higher excited states.
\item Requires all possile quadratic terms and they are all non-submodular.
\item Only reproduces the ground state manifold.
\end{itemize}

\examplesec
\begin{eqnarray}
-z_1z_2z_3z_4=-16\,b_1b_2b_3b_4+8\,(b_1b_2b_3+b_1b_2b_4+b_1b_3b_4+b_2b_3b_4)-\nonumber \\
4\,(b_1b_2+b_1b_3+b_1b_4+b_2b_3+b_2b_4+b_3b_4)+2\,(b_1+b_2+b_3+b_4)-1 \nonumber \\
\rightarrow 16\,t_a^2+8\,t_a\sum_{i=1}^4 b_i+8\,\sum_{i=1}^4\sum_{j>i}^4b_i\,b_j+16
\end{eqnarray}


\refsec
\begin{itemize}
\item Original paper: \cite{Rocchetto2016}.
\end{itemize}

\newpage

\subsection{PTR-CZW (Chancellor, Zohren, Warburton, 2017)}

\summarysec

Auxilliary qubits can be made to ``count'' the number of logical qubits in the $1$ configuration. By applying single qubit terms to the auxilliary qubits, the spectrum of \emph{any} permutation symmetric objective function can be reproduced. 

\costsec
\begin{itemize}
\item For a $k$ local coupler requires $k$ auxilliary qubits.
\end{itemize}

\prossec
\begin{itemize}
\item Natural flux qubit implementation \cite{Chancellor2017}.
\item Single gadget can reproduce any permutation symmetric spectrum.
\item High degree of symmetry means this method is natural for some kinds of quantum simulations \cite{Chancellor2016a}.
\end{itemize}

\conssec
\begin{itemize}
\item Requires coupling between all logical bits and from all logical bits to all auxilliary bits.
\item Requires single body terms of increasing strength as $k$ is increased.
\end{itemize}

\examplesec

A  $4$ qubit gadget guarantees that the number of auxillary bits in the $-1$ state is equal to the number of logical bits in the $1$ state
%
%
\begin{equation}
H_{4-\rm{count}}=  4\,\sum_{i=2}^4\sum_{j=1}^{i-1}b_ib_j+4\,\sum_{i=1}^4\sum_{j=1}^4b_ib_{a_j}-15\,\sum_{i=1}^4b_i-8\,\sum_{i=1}^4b_{a_i}+(5\,b_{a_1}+b_{a_2}-3\,b_{a_3}-7\,b_{a_4})+26
\end{equation}
This gadget can be expressed more naturally in terms of $z$:
\begin{equation}
H_{4-\rm{count}}=  \sum_{i=2}^4 \sum_{j =1}^{i-1} z_i z_j -\frac{1}{2} \sum_{i=1}^4  z_i +  \sum_{i=1}^4 \sum_{j=1}^4 z_i z_{a_j}  +\frac{1}{2}\left(5z_{a_1}+ z_{a_2}-3\,z_{a_3}-7 z_{a_4}\right).
\end{equation}
To replicate the spectrum of $b_1b_2b_3b_4$, we add
\begin{equation}
H_{2-\rm{local}}=-b_{a_4}+ \lambda H_{4-\rm{count}}.
\end{equation}
where $\lambda$ is a large number.

For the spectrum of 
\begin{align}
z_1z_2z_3z_4=16\,b_1b_2b_3b_4-8\,(b_1b_2b_3+b_1b_2b_4+b_1b_3b_4+b_2b_3b_4)+\nonumber \\
4\,(b_1b_2+b_1b_3+b_1b_4+b_2b_3+b_2b_4+b_3b_4)-2\,(b_1+b_2+b_3+b_4)+1,
\end{align}
 we implement,
\begin{equation}
H_{2-\rm{local}}=2\,b_{a_1}-2\,b_{a_2}+2\,b_{a_3}-2\,b_{a_4}+ \lambda H_{4-\rm{count}},
\end{equation}

\refsec
\begin{itemize}
\item Paper on flux qubit implementation: \cite{Chancellor2016}
\item Paper on MAX-$k$-SAT mapping: \cite{Chancellor2016} (published in a journal earlier than \cite{Chancellor2017} but put on arXiv 1 month later).
\item Talk including use in quantum simulation: \cite{Chancellor2016a}
\end{itemize}

\newpage

\subsection{Bit flipping (Ishikawa, 2011)}

\summarysec

For any variable $b$, we can consider the negation $\bar{b}=1-b$.
The process of exchanging $b$ for $\bar{b}$ is called \emph{flipping}.
Using bit-flipping, an arbitrary function in $n$ variables can be represented using at most $2^{(n-2)}(n-3)+1$ variables, though this is a gross overestimate.

Can be used in many different ways:
\begin{enumerate}
\item Flipping positive terms and using \ref{subsec:Negative-Monomial-Reduction}, recursively;
\item For $\alpha<0$, we can reduce $\alpha\bar{b}_{1}\bar{b}_{2}...\bar{b}_{k}$ very efficiently to submodular form using \ref{subsec:Negative-Monomial-Reduction}.
A generalized version exists for arbitrary combinations of flips in the monomial which makes reduction entirely submodular \cite{Ishikawa2011};
\item When we have quadratized we can minimize the number of non-submodular terms by flipping.
\item We can make use of both $b_{i}$ and $\bar{b}_{i}$ in the same objective function by adding on a sufficiently large penalty term: $\lambda(b_{i}+\bar{b}_{i}-1)^{2}=\lambda(1+2b_{i}\bar{b}_{i}-b_{i}-\bar{b}_{i})$.
This is similar to the ideas in reduction by substitution or deduc-reduc.
In this way, given a quadratic in $n$ variables we can make sure it only has at most $n$ nonsubmodular terms if we are willing to use the extra $n$ negation variables as well (so we have $2n$ variables in total).
\end{enumerate}

\costsec
\begin{itemize}
\item None, as replacing $b_{i}$ with it's negation $\bar{b}_{i}$ costs nothing except a trivial symbolic expansion.
\end{itemize}

\prossec
\begin{itemize}
\item Cheap and effective way of improving submodularity.
\item Can be used to combine terms in clever ways, making other methods more efficient.
\end{itemize}

\conssec
\begin{itemize}
\item Unless the form of the objective function is known, spotting these 'factorizations' using negations is difficult.
\item We need an auxiliary variable for each $b_i$ for which we also want to use $\bar{b_i}$ in the same objective function.
\end{itemize}

\examplesec

By bit-flipping $b_2$ and $b_4$, i.e. substituting $b_2 = 1 - \bar{b}_{2}$ and $b_4 = 1 - \bar{b}_{4}$, we see that:
\begin{eqnarray}
H & = & 3b_{1}b_{2}+b_{2}b_{3}+2b_{1}b_{4}-4b_{2}b_{4}\protect\\
 & = & -3b_{1}\bar{b}_{2}-\bar{b}_{2}b_{3}-2b_{1}\bar{b}_{4}-\bar{b}_{2}\bar{b}_{4}+5b_{1}+b_{3}+4\bar{b}_{2}+4\bar{b}_{4}-4.
\end{eqnarray}
The first expression is highly non-submodular while the second is entirely submodular.

\refsec
\begin{itemize}
\item Original paper: \cite{Ishikawa2011}.
\end{itemize}

\newpage

\section{Methods that quadratize MULTIPLE terms with the SAME auxiliaries (Case 1: Symmetric Function Reductions, SFR)}

A symmetric function is one where if we switch any of the variable names (for example $b_1 \rightarrow b_5 \rightarrow b_8 \rightarrow b_1$), the function's output is unaffected. 

\subsection{SFR-ABCG-1 (Anthony, Boros, Crama, Gruber, 2014)}

\summarysec

Any $n$-variable symmetric function $f\left(b_1,b_2,\ldots b_n\right)\equiv f(b)$  can be quadratized with $n-2$ auxiliaries:

\begin{align}
f(b)&\rightarrow -\alpha_{0}-\alpha_{0}\sum_{i}b_{i}+a_{2}\sum_{ij}b_{i}b_{j}+2\sum_{i}\left(\alpha_{i}-c\right)b_{a_{i}}\left(2i-\frac{1}{2}-\sum_{j}b_{j}\right) \\
  c  &= \begin{cases}
{\rm min}\left(\alpha_{2j}\right) & ,i\in\text{even}\\
{\rm min}\left(\alpha_{2j-1}\right) & ,i\in\text{odd}
\end{cases} \\
a_2 &= \textrm{Determined from Page 12 of \cite{Anthony2015}}
\end{align}

\costsec
\begin{itemize}
\item $n-2$ auxiliaries for any $n$-variable symmetric function.
\item $n^2$ non-submodular quadratic terms (of the non-auxiliary variables).
\item $n-2$ non-submodular linear terms (of the auxiliary variables).
\end{itemize}

\prossec
\begin{itemize}
\item Quadratization is symmetric in all non-auxiliary variables (this is not always true, for example some of the methods in the  NTR section).
\item Reproduces the full spectrum.
\item When there's a large number of terms, there's fewer auxiliary variables than quadratizing each positive monomial separately.
\end{itemize}

\conssec
\begin{itemize}
\item Only works on a specific class of functions, although the quadratizations of arbitrary functions can be related to the quadratizations of symmetric functions on a larger number of variables. 
\item All quadratic terms of the non-auxiliary variables are non-sub-modular.
\item All linear terms of the auxliaries are non-submodular.
\item Not meant so much to be practical, but rather an easy proof of an upper bound on the number of needed auxiliaries.
\end{itemize}

\refsec
\begin{itemize}
\item 2014, original paper (Theorem 4.1, with $\alpha_i$ from Corollary 2.3): \cite{Anthony2014}.
\end{itemize}

\newpage

\subsection{SFR-BCR-1 (Boros, Crama, Rodr\'{i}guez-Heck, 2018)}

\summarysec

Any $n$-variable symmetric function $f\left(b_1,b_2,\ldots b_n\right)\equiv f(b)$  that is non-zero only when $\sum b_i = c$ where $\nicefrac{n}{2}\le c \le n$, can be quadratized with $m=\lceil \textrm{log}_2c \rceil +1$ auxiliary variables:

{\scriptsize
\begin{align}
\hspace{-5mm}
f(b) \rightarrow       
\alpha
+\alpha^{b}            \hspace{-0.5mm}    \sum_{i}                                b_{i}
+\alpha_1^{b_{a}}      \hspace{-1.0mm}    \sum_{i}^{m-1}                          b_{a_{i}}
+\alpha_2^{b_{a}}                                                                 b_{a_{m}}
+\alpha^{bb}           \hspace{-0.5mm}    \sum_{ij}                               b_{i}b_{j}
+\alpha_1^{bb_{a}}     \hspace{-0.5mm}    \sum_{i} \hspace{-1.5mm}\sum_{j}^{m-1}  b_{i}b_{a_{j}}
+\alpha_2^{bb_{a}}     \hspace{-0.5mm}    \sum_{i}b_{i}                           b_{a_{m}}
+\alpha_1^{b_{a}b_{a}} \hspace{-1.2mm}    \sum_{ij}^{m-1}                         b_{a_{i}}b_{a_{j}} 
+\alpha_2^{b_{a}b_{a}} \hspace{-1.2mm}    \sum_{i}^{m-1}                          b_{a_{i}}b_{a_{m}},
\end{align}
}

\noindent where:

\begin{align}
\begin{pmatrix}
\alpha           & \alpha^{bb}\\
\alpha^{b}       & \alpha_1^{bb_{a}}\\
\alpha_1^{b_{a}} & \alpha_2^{bb_{a}}\\
\alpha_2^{b_{a}} & \alpha_1^{b_{a}b_{a}}\\
\cdot            & \alpha_2^{b_{a}b_{a}}
\end{pmatrix}=\begin{pmatrix}(c+1)^{2} & 1\\
-2(c+1) & -2^{i}\\
(c+1)2^{i} & 2\left(1+2^{m-1}\right)\\
\left(1+2^{m-1}\right)\left(2^{m-1}-2c-1\right) & 2^{i+j-1}\\
\cdot  & - \left(1+2^{m-1}\right)2^{i}
\end{pmatrix}.
\end{align}

\costsec
\begin{itemize}
\item $m=\lceil \textrm{log}_2 c \rceil + 1$ auxiliary variables. 
\item $n^2 + m^2$ non-submodular quadratic terms (all possible quadratic terms involving only non-auxiliary or only auxiliary variables).
\item $m$ non-submodular linear terms (all possible linear terms involving auxiliaries).
\end{itemize}

\prossec
\begin{itemize}
\item Small number of auxiliary terms
\end{itemize}

\conssec
\begin{itemize}
\item Only works for a special class of functions
\item Introduces many linear and even more quadratic non-submodular terms.
\end{itemize}

\examplesec

For $n = 4$ and $c = 2$, we have $m = 2$, and 
\begin{eqnarray}
	& &\hspace{-10mm}(b_1 b_2 +b_1 b_3 + b_1 b_4 + b_2 b_3 + b_2 b_4 + b_3 b_4)  
	- 3(b_1 b_2 b_3 + b_1 b_2 b_4 + b_1 b_3 b_4 + b_2 b_3 b_4) + 6 b_1 b_2 b_3 b_4 \\
	&\to&  \left(-3 + b_1+b_2 + b_3 + b_4 - b_{a_1} + 3 b_{a_2}\right)^2
\end{eqnarray}
using the alternate form below.

\altformsec
\begin{align}
f\left(b_{1},b_{2},\ldots,b_{n}\right) &\rightarrow \left(-(c+1)+\sum_{i}b_{i}-\sum_{i}^{m-1}2^{i-1}b_{a_{i}}+\left(1+2^{m-1}\right)b_{a_{m}}\right)^{2}
\end{align}

\refsec
\begin{itemize}
\item 2018, original paper (Theorem 1): \cite{Boros2018QuadratizationsOS}.
\end{itemize}

\newpage

\subsection{SFR-BCR-2 (Boros, Crama, Rodr\'{i}guez-Heck, 2018)}

\summarysec

Any $n$-variable symmetric function $f\left(b_1,b_2,\ldots b_n\right)\equiv f(b)$  that is non-zero only when $\sum b_i = c$ where $0\le c \le\nicefrac{n}{2} $, can be quadratized with $m=\lceil \textrm{log}_2(n-c) \rceil +1$ auxiliary variables:

{\scriptsize
\begin{align}
\hspace{-5mm}
f(b) \rightarrow       
\alpha
+\alpha^{b}            \hspace{-0.5mm}    \sum_{i}                                b_{i}
+\alpha_1^{b_{a}}      \hspace{-1.0mm}    \sum_{i}^{m-1}                          b_{a_{i}}
+\alpha_2^{b_{a}}                                                                 b_{a_{m}}
+\alpha^{bb}           \hspace{-0.5mm}    \sum_{ij}                               b_{i}b_{j}
+\alpha_1^{bb_{a}}     \hspace{-0.5mm}    \sum_{i} \hspace{-1.5mm}\sum_{j}^{m-1}  b_{i}b_{a_{j}}
+\alpha_2^{bb_{a}}     \hspace{-0.5mm}    \sum_{i}b_{i}                           b_{a_{m}}
+\alpha_1^{b_{a}b_{a}} \hspace{-1.2mm}    \sum_{ij}^{m-1}                         b_{a_{i}}b_{a_{j}} 
+\alpha_2^{b_{a}b_{a}} \hspace{-1.2mm}    \sum_{i}^{m-1}                          b_{a_{i}}b_{a_{m}},
\end{align}
}

\noindent where:

\begin{align}
\begin{pmatrix}\alpha & \alpha^{bb}\\
\alpha^{b} & \alpha_1^{bb_{a}}\\
\alpha_1^{b_{a}} & \alpha_2^{bb_{a}}\\
\alpha_2^{b_{a}} & \alpha_1^{b_{a}b_{a}}\\
\cdot  & \alpha_2^{b_{a}b_{a}}
\end{pmatrix}=\begin{pmatrix}(c-1)^{2} & 1\\
-2(c-1) & +2^{i}\\
(1-c)2^{i} &-2\left(1+2^{m-1}\right)\\
\left(1+2^{m-1}\right)\left(2^{m-1}+2c-1\right) & 2^{i+j-1}\\
\cdot  & - \left(1+2^{m-1}\right)2^{i}
\end{pmatrix}.
\end{align}

\costsec
\begin{itemize}
\item $m=\lceil \textrm{log}_2 (n-c) \rceil + 1$ auxiliary variables.
\item $n^2 + m^2$ non-submodular quadratic terms (all possible quadratic terms involving only non-auxiliary or only auxiliary variables).
\item $m$ non-submodular linear terms (all possible linear terms involving auxiliaries).
\end{itemize}

\prossec
\begin{itemize}
\item Small number of auxiliary terms
\end{itemize}

\conssec
\begin{itemize}
\item Only works for a special class of functions
\item Introduces many linear and even more quadratic non-submodular terms.
\end{itemize}

\examplesec

For $n = 4$ and $c = 2$, we have $m = 2$, and 
\begin{eqnarray}
	& &\hspace{-10mm}(b_1 b_2 +b_1 b_3 + b_1 b_4 + b_2 b_3 + b_2 b_4 + b_3 b_4)  
	- 3(b_1 b_2 b_3 + b_1 b_2 b_4 + b_1 b_3 b_4 + b_2 b_3 b_4) + 6 b_1 b_2 b_3 b_4 \\
	&\to&  \left(1 - b_1 -b_2 - b_3 - b_4 - b_{a_1} + 3 b_{a_2}\right)^2
\end{eqnarray}
using the alternate form below.

\altformsec
\begin{align}
f\left(b_{1},b_{2},\ldots,b_{n}\right) &\rightarrow \left((c-1)-\sum_{i}b_{i}-\sum_{i}^{m-1}2^{i-1}b_{a_{i}}+\left(1+2^{m-1}\right)b_{a_{m}}\right)^{2}
\end{align}

\refsec
\begin{itemize}
\item 2018, original paper (Theorem 1): \cite{Boros2018QuadratizationsOS}.
\end{itemize}

\newpage

\subsection{SFR-BCR-3 (Boros, Crama, Rodr\'{i}guez-Heck, 2018)}

\summarysec

Any $n$-variable symmetric function $f\left(b_1,b_2,\ldots b_n\right)\equiv f(b)$  that is non-zero only when $\sum b_i = c$ where $\nicefrac{n}{2}\le c \le n$, can be quadratized with $m=\lceil \textrm{log}_2c \rceil $ auxiliary variables $f\rightarrow$:

{\scriptsize
\begin{align}
\hspace{-5mm}
f(b) \rightarrow       
\alpha
+\alpha^{b}            \hspace{-0.5mm}    \sum_{i}                                b_{i}
+\alpha_1^{b_{a}}      \hspace{-1.0mm}    \sum_{i}^{m-1}                          b_{a_{i}}
+\alpha_2^{b_{a}}                                                                 b_{a_{m}}
+\alpha^{bb}           \hspace{-0.5mm}    \sum_{ij}                               b_{i}b_{j}
+\alpha_1^{bb_{a}}     \hspace{-0.5mm}    \sum_{i} \hspace{-1.5mm}\sum_{j}^{m-1}  b_{i}b_{a_{j}}
+\alpha_2^{bb_{a}}     \hspace{-0.5mm}    \sum_{i}b_{i}                           b_{a_{m}}
+\alpha_1^{b_{a}b_{a}} \hspace{-1.2mm}    \sum_{ij}^{m-1}                         b_{a_{i}}b_{a_{j}} 
+\alpha_2^{b_{a}b_{a}} \hspace{-1.2mm}    \sum_{i}^{m-1}                          b_{a_{i}}b_{a_{m}},
\end{align}
}

\noindent where:

\begin{align}
\begin{pmatrix}
\alpha           & \alpha^{bb}\\
\alpha^{b}       & \alpha_1^{bb_{a}}\\
\alpha_1^{b_{a}} & \alpha_2^{bb_{a}}\\
\alpha_2^{b_{a}} & \alpha_1^{b_{a}b_{a}}\\
\cdot                 & \alpha_2^{b_{a}b_{a}}
\end{pmatrix}=\begin{pmatrix}\frac{1}{2}(c^2+3c+2) & \frac{1}{2}\\
-c-\frac{3}{2} & -2^{i}\\
(3+c)2^{i-1} & \left(1+2^{m}\right)\\
\left(1+2^{m}\right)\left(2^{m-1}-c-1\right) & 2^{i+j-1}\\
\cdot & - \left(1+2^{m}\right)2^{i}
\end{pmatrix}.
\end{align}

\costsec
\begin{itemize}
\item $m=\lceil \textrm{log}_2 c \rceil$ auxiliary variables. 
\item $n^2 + m^2$ non-submodular quadratic terms (all possible quadratic terms involving only non-auxiliary or only auxiliary variables).
\item $m$ non-submodular linear terms (all possible linear terms involving auxiliaries).
\end{itemize}

\prossec
\begin{itemize}
\item Small number of auxiliary terms
\end{itemize}

\conssec
\begin{itemize}
\item Only works for a special class of functions
\item Introduces many linear and even more quadratic non-submodular terms.
\end{itemize}

\examplesec

For $n = 4$ and $c = 2$, we have $m = 1$, and 
\begin{eqnarray}
	& &\hspace{-15mm}(b_1 b_2 +b_1 b_3 + b_1 b_4 + b_2 b_3 + b_2 b_4 + b_3 b_4)  
	- 3(b_1 b_2 b_3 + b_1 b_2 b_4 + b_1 b_3 b_4 + b_2 b_3 b_4) + 6 b_1 b_2 b_3 b_4 \\
	&\to&  \binom{-3+b_1 + b_2 + b_3 + b_4 + 3b_{a_1}}{2}
\end{eqnarray}
using the alternate form below.

\altformsec
\begin{align}
f\left(b_{1},b_{2},\ldots,b_{n}\right)&\rightarrow \binom{-(c+1)+\sum_{i}b_{i}-\sum_{i}^{m-1}2^{i}b_{a_{i}}+\left(1+2^{m}\right)b_{a_{m}}}{2}
\end{align}

\refsec
\begin{itemize}
\item 2018, original paper (Theorem 2): \cite{Boros2018QuadratizationsOS}.
\end{itemize}

\newpage

\subsection{SFR-BCR-4 (Boros, Crama, Rodr\'{i}guez-Heck, 2018)}

\summarysec

Any $n$-variable symmetric function $f\left(b_1,b_2,\ldots b_n\right)\equiv f(b)$  that is non-zero only when $\sum b_i = c$ where $0\le c \le\nicefrac{n}{2} $, can be quadratized with $m=\lceil \textrm{log}_2(n-c) \rceil $ auxiliary variables $f\rightarrow$:

{\scriptsize
\begin{align}
\hspace{-5mm}
f(b) \rightarrow       
\alpha
+\alpha^{b}            \hspace{-0.5mm}    \sum_{i}                                b_{i}
+\alpha_1^{b_{a}}      \hspace{-1.0mm}    \sum_{i}^{m-1}                          b_{a_{i}}
+\alpha_2^{b_{a}}                                                                 b_{a_{m}}
+\alpha^{bb}           \hspace{-0.5mm}    \sum_{ij}                               b_{i}b_{j}
+\alpha_1^{bb_{a}}     \hspace{-0.5mm}    \sum_{i} \hspace{-1.5mm}\sum_{j}^{m-1}  b_{i}b_{a_{j}}
+\alpha_2^{bb_{a}}     \hspace{-0.5mm}    \sum_{i}b_{i}                           b_{a_{m}}
+\alpha_1^{b_{a}b_{a}} \hspace{-1.2mm}    \sum_{ij}^{m-1}                         b_{a_{i}}b_{a_{j}} 
+\alpha_2^{b_{a}b_{a}} \hspace{-1.2mm}    \sum_{i}^{m-1}                          b_{a_{i}}b_{a_{m}},
\end{align}
}

\noindent where:

\begin{align}
\begin{pmatrix}
\alpha           & \alpha^{bb}\\
\alpha^{b}       & \alpha_1^{bb_{a}}\\
\alpha_1^{b_{a}} & \alpha_2^{bb_{a}}\\
\alpha_2^{b_{a}} & \alpha_1^{b_{a}b_{a}}\\
\cdot            & \alpha_2^{b_{a}b_{a}}
\end{pmatrix}=\begin{pmatrix}\frac{1}{2}(c^2-3c+2) & \frac{1}{2}\\
-c+\frac{3}{2} & +2^{i}\\
(3-c)2^{i-1} &-\left(1+2^{m}\right)\\
\left(1+2^{m}\right)\left(2^{m-1}+c-1\right) & 2^{i+j-1}\\
\cdot  & - \left(1+2^{m}\right)2^{i}
\end{pmatrix}.
\end{align}

\costsec
\begin{itemize}
\item $m=\lceil \textrm{log}_2 (n-c) \rceil$ auxiliary variables.
\item $n^2 + m^2$ non-submodular quadratic terms (all possible quadratic terms involving only non-auxiliary or only auxiliary variables).
\item $m$ non-submodular linear terms (all possible linear terms involving auxiliaries).
\end{itemize}

\prossec
\begin{itemize}
\item Small number of auxiliary terms
\end{itemize}

\conssec
\begin{itemize}
\item Only works for a special class of functions
\item Introduces many linear and even more quadratic non-submodular terms.
\end{itemize}

\examplesec

For $n = 4$ and $c = 2$, we have $m = 1$, and 
\begin{eqnarray}
	& &\hspace{-15mm}(b_1 b_2 +b_1 b_3 + b_1 b_4 + b_2 b_3 + b_2 b_4 + b_3 b_4)  
	- 3(b_1 b_2 b_3 + b_1 b_2 b_4 + b_1 b_3 b_4 + b_2 b_3 b_4) + 6 b_1 b_2 b_3 b_4 \\
	&\to&  \binom{1 - b_1 - b_2 - b_3 - b_4 + 3b_{a_1}}{2}
\end{eqnarray}
using the alternate form below.

\altformsec
\begin{align}
f\left(b_{1},b_{2},\ldots,b_{n}\right)& \rightarrow \binom{(c-1)-\sum_{i}b_{i}-\sum_{i}^{m-1}2^{i}b_{a_{i}}+\left(1+2^{m}\right)b_{a_{m}}}{2}
\end{align}

\refsec
\begin{itemize}
\item 2018, original paper (Theorem 2): \cite{Boros2018QuadratizationsOS}.
\end{itemize}

\newpage

\subsection{SFR-BCR-5 (Boros, Crama, Rodr\'{i}guez-Heck, 2018)}

\summarysec

For an $n$-variable symmetric function that is a function of the sum of all variables $f\left(b_1,b_2,\ldots b_n\right) =  f(\sum b_i)$, for some large value of $\lambda > \textrm{max}(f)$, and $\lceil \sqrt{n+1}\rceil$:

{\footnotesize
\begin{align}
\begin{gathered}
f\left(\sum b_i \right) \rightarrow  \sum_{ij}^{m}f\left((i-1)\left(m+1\right)+(j-1)\right)b_{a_{i}}b_{a_{c+j}}+\lambda\left(\left(1-\sum_{i}^{m}b_{a_{i}}\right)^{2}+\left(1-\sum_{i}^{m}b_{a_{c+i}}\right)^{2}+\right.\\
\left.\left(\sum_{i}b_{i}-\left(\left(m+1\right)\sum_{i}^{m}(i-1)y_{a_{i}}+\sum_{i}^{m}(i-1)b_{a_{c+i}}\right)\right)^{2}+\left(\sum_{i}b_{i}-\left(\left(m+1\right)\sum_{i}^{m}(i-1)y_{a_{i}}+\sum_{i}^{m}(i-1)b_{a_{c+i}}\right)\right)^{2}\right)
\end{gathered}
\end{align}
}

\noindent where:

\begin{align}
\begin{pmatrix}\alpha & \alpha^{bb}\\
\alpha^{b} & \alpha^{bb_{a,1}}\\
\alpha^{b_{a,1}} & \alpha^{bb_{a,2}}\\
\alpha^{b_{a,2}} & \alpha^{b_{a}b_{a}}
\end{pmatrix}=\begin{pmatrix}(c+1)^{2} & 1\\
-2(c+1) & -2^{i}\\
2(c+1) & -2\left(2m-2^{m-1}+1\right)\\
2(c+1)\left(2c-2^{m-1}+1\right) & 2^{i+j-2}
\end{pmatrix}.
\end{align}

\costsec
\begin{itemize}
\item $m=\lceil \sqrt{n+1} \rceil $ auxiliary variables.
\item $n^2 + m^2$ non-submodular quadratic terms (all possible quadratic terms involving only non-auxiliary or only auxiliary variables).
\item $m$ non-submodular linear terms (all possible linear terms involving auxiliaries).
\end{itemize}

\prossec
\begin{itemize}
\item Small number of auxiliary terms
\end{itemize}

\conssec
\begin{itemize}
\item Only works for a special class of functions
\item Introduces many linear and even more quadratic non-submodular terms.
\end{itemize}

\examplesec

By bit-flipping $b_2$ and $b_4$, i.e. substituting $b_2 = 1 - \bar{b}_{2}$ and $b_4 = 1 - \bar{b}_{4}$, we see that:
\begin{eqnarray}
H & = & 3b_{1}b_{2}+b_{2}b_{3}+2b_{1}b_{4}-4b_{2}b_{4}\protect\\
 & = & -3b_{1}\bar{b}_{2}-\bar{b}_{2}b_{3}-2b_{1}\bar{b}_{4}-\bar{b}_{2}\bar{b}_{4}+5b_{1}+b_{3}+4\bar{b}_{2}+4\bar{b}_{4}-4.
\end{eqnarray}
The first expression is highly non-submodular while the second is entirely submodular.

\altformsec
\begin{align}
f\left(b_{1},b_{2},\ldots,b_{n}\right)& \rightarrow \binom{\frac{1}{2}\left(\left(c-1\right)-\sum_{i}b_{i}-\left(2(n-c)-2^{m-1}+1\right)b_{a_{m}}-\sum_{i}^{m-1}2^{i-1}b_{a_{i}}\right)}{2}
\end{align}

\refsec
\begin{itemize}
\item 2018, original paper (Theorem 9): \cite{Boros2018QuadratizationsOS} (contains a typo which was corrected in Theorem 6 of \cite{Boros2018boundsPaper}.
\end{itemize}

\newpage

\subsection{SFR-BCR-6 (Boros, Crama, Rodr\'{i}guez-Heck, 2018)}

\summarysec

For an $n$-variable symmetric function that is a function of a \textit{weighted} sum of all variables $f\left(b_1,b_2,\ldots b_n\right) =  f(\sum w_ib_i)$, for some large value of $\lambda > \textrm{max}(f)$, and $\max \left( f\left(\sum w_i b_i \right)\right) < (m+1)^2$:

\begin{align}
\begin{gathered}
f\left(\sum w_{i}b_{i}\right)\rightarrow\sum_{ij}^{m}\alpha_{ij}b_{a_{i}}b_{a_{m+i}}+\lambda\left(1+\left(\sum_{i}w_{i}b_{i}-(m-1)\sum_{i}^{m}b_{a_{i}}+\sum_{i}^{m}b_{a_{c+i}}\right)^{2}\right.\\
+\left.\sum_{i}^{m-1}\left(1-b_{a_{i}}\right)b_{a_{i+1}}+\sum_{i}^{m-1}\left(1-b_{a_{i+m}}\right)b_{a_{i+m+1}}\right)
\end{gathered}
\end{align}

\noindent where:

\begin{align}
\sum_{i}^{\alpha}\sum_{j}^{\beta}\alpha_{ij}&=f\left(\alpha(m+1)+\beta\right)
\end{align}

\costsec
\begin{itemize}
\item $2m$ auxiliary variables, where $m> \sqrt{\max \left( f \left( \sum w_i b_i \right)  \right)}-1$.
\item $n^2 + m^2$ non-submodular quadratic terms (all possible quadratic terms involving only non-auxiliary or only auxiliary variables).
\item $m$ non-submodular linear terms (all possible linear terms involving auxiliaries).
\end{itemize}

\prossec
\begin{itemize}
\item Small number of auxiliary terms
\end{itemize}

\conssec
\begin{itemize}
\item Only works for a special class of functions
\item Introduces many linear and even more quadratic non-submodular terms.
\end{itemize}

\examplesec

By bit-flipping $b_2$ and $b_4$, i.e. substituting $b_2 = 1 - \bar{b}_{2}$ and $b_4 = 1 - \bar{b}_{4}$, we see that:
\begin{eqnarray}
H & = & 3b_{1}b_{2}+b_{2}b_{3}+2b_{1}b_{4}-4b_{2}b_{4}\protect\\
 & = & -3b_{1}\bar{b}_{2}-\bar{b}_{2}b_{3}-2b_{1}\bar{b}_{4}-\bar{b}_{2}\bar{b}_{4}+5b_{1}+b_{3}+4\bar{b}_{2}+4\bar{b}_{4}-4.
\end{eqnarray}
The first expression is highly non-submodular while the second is entirely submodular.

\altformsec
\begin{align}
f\left(b_{1},b_{2},\ldots,b_{n}\right)& \rightarrow \binom{\frac{1}{2}\left(\left(c-1\right)-\sum_{i}b_{i}-\left(2(n-c)-2^{m-1}+1\right)b_{a_{m}}-\sum_{i}^{m-1}2^{i-1}b_{a_{i}}\right)}{2}
\end{align}

\refsec
\begin{itemize}
\item 2018, original paper (Theorem 10): \cite{Boros2018QuadratizationsOS}.
\end{itemize}

\newpage


\subsection{SFR-ABCG-2 (Anthony, Boros, Crama, Gruber, 2014)}

\summarysec

For any $n$-variable, $k$-local function that is non-zero  only if $\sum b_{i}=2m-1$, we call it the "partity function" and it can be quadratized as follows:

\begin{align}
f\left(b_{1},b_{2},\ldots,b_{n}\right)\rightarrow\sum_{i}b_{i}+2\sum_{ij}b_{i}b_{j}+4\sum_{2i-1}^{n-1}b_{a_{i}}\left(2i-1-\sum_{j}b_{j}\right).
\end{align}

\costsec
\begin{itemize}
\item $m=2\lfloor n/2 \rfloor$ auxiliary variables.
\item $\lfloor 1.5n\rfloor$ non-submodular linear terms.
\item $ n^2$ non-submodular quadratic terms.
\end{itemize}

\prossec
\begin{itemize}
\item Smaller number of auxiliary variables than the most naive methods.
\end{itemize}

\conssec
\begin{itemize}
\item Only works for a special class of functions
\item Introduces many linear and even more quadratic non-submodular terms (everything is non-submodular except for $0.5n^2$ quadratic terms involving the auxiliaries with the non-auxiliaries).
\item non-submodular terms can be rather large compared to the submodular terms (about $4n$ times as big).
\end{itemize}

\examplesec

By bit-flipping $b_2$ and $b_4$, i.e. substituting $b_2 = 1 - \bar{b}_{2}$ and $b_4 = 1 - \bar{b}_{4}$, we see that:
\begin{eqnarray}
H & = & 3b_{1}b_{2}+b_{2}b_{3}+2b_{1}b_{4}-4b_{2}b_{4}\protect\\
 & = & -3b_{1}\bar{b}_{2}-\bar{b}_{2}b_{3}-2b_{1}\bar{b}_{4}-\bar{b}_{2}\bar{b}_{4}+5b_{1}+b_{3}+4\bar{b}_{2}+4\bar{b}_{4}-4.
\end{eqnarray}
The first expression is highly non-submodular while the second is entirely submodular.


\refsec
\begin{itemize}
\item 2014, original paper (Theorem 4.6): \cite{Anthony2014,Anthony2016}.  
\end{itemize}

\newpage

\subsection{SFR-ABCG-3 (Anthony, Boros, Crama, Gruber, 2014)}

\summarysec

The complement of the parity function can be quadratized as follow:

\begin{align}
f\left(b_{1},b_{2},\ldots,b_{n}\right)\rightarrow1+2\sum_{ij}b_{i}b_{j}-\sum_{i}b_{i}+4\sum_{2i}^{n-1}b_{a_{i}}\left(i-\sum_{j}^{n}b_{j}\right)
\end{align}

\costsec
\begin{itemize}
\item $m=2\lfloor \frac{n-1}{2} \rfloor$ auxiliary variables.
\item $\lfloor 0.5n\rfloor$ non-submodular linear terms.
\item $ n^2$ non-submodular quadratic terms.
\end{itemize}

\prossec
\begin{itemize}
\item Smaller number of auxiliary variables than the most naive methods.
\item Fewer non-submodular linear terms than in the analogous quadratization for its complement (the parity function).
\end{itemize}

\conssec
\begin{itemize}
\item Only works for a special class of functions
\item Introduces many linear and even more quadratic non-submodular terms (everything is non-submodular except for $0.5n^2$ quadratic terms involving the auxiliaries with the non-auxiliaries, and all $n$ linear terms involving only the non-auxiliaries).
\item non-submodular terms can be rather large compared to the submodular terms (about $4n$ times as big).
\end{itemize}

\examplesec

By bit-flipping $b_2$ and $b_4$, i.e. substituting $b_2 = 1 - \bar{b}_{2}$ and $b_4 = 1 - \bar{b}_{4}$, we see that:
\begin{eqnarray}
H & = & 3b_{1}b_{2}+b_{2}b_{3}+2b_{1}b_{4}-4b_{2}b_{4}\protect\\
 & = & -3b_{1}\bar{b}_{2}-\bar{b}_{2}b_{3}-2b_{1}\bar{b}_{4}-\bar{b}_{2}\bar{b}_{4}+5b_{1}+b_{3}+4\bar{b}_{2}+4\bar{b}_{4}-4.
\end{eqnarray}
The first expression is highly non-submodular while the second is entirely submodular.


\refsec
\begin{itemize}
\item 2014, original paper (Theorem 4.6): \cite{Anthony2014,Anthony2016}.
\end{itemize}

\newpage

\subsection{SFR-BCR-7 (Boros, Crama,  Rodr\'{i}guez-Heck, 2018)}


\summarysec

For a symmetric function such that $f\left(|c|\right)=0$ for $c>n$, then with with $2\lceil \sqrt{n+1}\rceil$ auxiliary variables, we have:

\begin{align}
f\left(b_{1},b_{2},\ldots,b_{n}\right)\rightarrow1+2\sum_{ij}b_{i}b_{j}-\sum_{i}b_{i}+4\sum_{2i}^{n-1}b_{a_{i}}\left(i-\sum_{j}^{n}b_{j}\right)
\end{align}

\costsec
\begin{itemize}
\item $m=\lceil \sqrt{n+1}\rceil$ auxiliary variables.
\item $\lfloor 0.5n\rfloor$ non-submodular linear terms. 
\item $ n^2$ non-submodular quadratic terms. 
\end{itemize}

\prossec
\begin{itemize}
\item Smaller number of auxiliary variables than the most naive methods. 
\item Fewer non-submodular linear terms than in the analogous quadratization for its complement (the parity function). 
\end{itemize}

\conssec
\begin{itemize}
\item Only works for a special class of functions
\item Introduces many linear and even more quadratic non-submodular terms (everything is non-submodular except for $0.5n^2$ quadratic terms involving the auxiliaries with the non-auxiliaries, and all $n$ linear terms involving only the non-auxiliaries).
\item non-submodular terms can be rather large compared to the submodular terms (about $4n$ times as big).
\end{itemize}

\examplesec

By bit-flipping $b_2$ and $b_4$, i.e. substituting $b_2 = 1 - \bar{b}_{2}$ and $b_4 = 1 - \bar{b}_{4}$, we see that:
\begin{eqnarray}
H & = & 3b_{1}b_{2}+b_{2}b_{3}+2b_{1}b_{4}-4b_{2}b_{4}\protect\\
 & = & -3b_{1}\bar{b}_{2}-\bar{b}_{2}b_{3}-2b_{1}\bar{b}_{4}-\bar{b}_{2}\bar{b}_{4}+5b_{1}+b_{3}+4\bar{b}_{2}+4\bar{b}_{4}-4.
\end{eqnarray}
The first expression is highly non-submodular while the second is entirely submodular.


\refsec
\begin{itemize}
\item 2018, original paper (Theorem 6): \cite{Boros2018boundsPaper}.
\end{itemize}

\newpage

\subsection{SFR-BCR-8 (Boros, Crama,  Rodr\'{i}guez-Heck, 2018)}

\summarysec

For a symmetric function such that $f\left(|c|\right)=0$ for $c>n$, then with with $\max\left( \lceil \log(c)\rceil , \lceil \log(n-c)\rceil  \right)$ auxiliary variables, we have:

\begin{align}
f\left(b_{1},b_{2},\ldots,b_{n}\right)\rightarrow1+2\sum_{ij}b_{i}b_{j}-\sum_{i}b_{i}+4\sum_{2i}^{n-1}b_{a_{i}}\left(i-\sum_{j}^{n}b_{j}\right)
\end{align}

\costsec
\begin{itemize}
\item $\max\left( \lceil \log(c)\rceil , \lceil \log(n-c)\rceil  \right)$ auxiliary variables.
\item $\lfloor 0.5n\rfloor$ non-submodular linear terms. 
\item $ n^2$ non-submodular quadratic terms. 
\end{itemize}

\prossec
\begin{itemize}
\item Smaller number of auxiliary variables than the most naive methods. 
\item Fewer non-submodular linear terms than in the analogous quadratization for its complement (the parity function). 
\end{itemize}

\conssec
\begin{itemize}
\item Only works for a special class of functions
\item Introduces many linear and even more quadratic non-submodular terms (everything is non-submodular except for $0.5n^2$ quadratic terms involving the auxiliaries with the non-auxiliaries, and all $n$ linear terms involving only the non-auxiliaries).
\item non-submodular terms can be rather large compared to the submodular terms (about $4n$ times as big).
\end{itemize}

\examplesec

By bit-flipping $b_2$ and $b_4$, i.e. substituting $b_2 = 1 - \bar{b}_{2}$ and $b_4 = 1 - \bar{b}_{4}$, we see that:
\begin{eqnarray}
H & = & 3b_{1}b_{2}+b_{2}b_{3}+2b_{1}b_{4}-4b_{2}b_{4}\protect\\
 & = & -3b_{1}\bar{b}_{2}-\bar{b}_{2}b_{3}-2b_{1}\bar{b}_{4}-\bar{b}_{2}\bar{b}_{4}+5b_{1}+b_{3}+4\bar{b}_{2}+4\bar{b}_{4}-4.
\end{eqnarray}
The first expression is highly non-submodular while the second is entirely submodular.


\refsec
\begin{itemize}
\item 2018, original paper (Theorem 7): \cite{Boros2018boundsPaper}.
\end{itemize}

\newpage

\subsection{SFR-BCR-9 (Boros, Crama,  Rodr\'{i}guez-Heck, 2018)}

\summarysec

For a symmetric function such that $f\left(|c|\right)=0$ for $c>n$, then with with $\max\left( \lceil \log(c)\rceil , \lceil \log(n-c)\rceil  \right)$ auxiliary variables, we have:

\begin{align}
f\left(b_{1},b_{2},\ldots,b_{n}\right)\rightarrow1+2\sum_{ij}b_{i}b_{j}-\sum_{i}b_{i}+4\sum_{2i}^{n-1}b_{a_{i}}\left(i-\sum_{j}^{n}b_{j}\right)
\end{align}

\costsec
\begin{itemize}
\item $\max\left( \lceil \log(c)\rceil , \lceil \log(n-c)\rceil  \right)$ auxiliary variables.
\item $\lfloor 0.5n\rfloor$ non-submodular linear terms. 
\item $ n^2$ non-submodular quadratic terms. 
\end{itemize}

\prossec
\begin{itemize}
\item Smaller number of auxiliary variables than the most naive methods. 
\item Fewer non-submodular linear terms than in the analogous quadratization for its complement (the parity function). 
\end{itemize}

\conssec
\begin{itemize}
\item Only works for a special class of functions
\item Introduces many linear and even more quadratic non-submodular terms (everything is non-submodular except for $0.5n^2$ quadratic terms involving the auxiliaries with the non-auxiliaries, and all $n$ linear terms involving only the non-auxiliaries).
\item non-submodular terms can be rather large compared to the submodular terms (about $4n$ times as big).
\end{itemize}

\examplesec

By bit-flipping $b_2$ and $b_4$, i.e. substituting $b_2 = 1 - \bar{b}_{2}$ and $b_4 = 1 - \bar{b}_{4}$, we see that:
\begin{eqnarray}
H & = & 3b_{1}b_{2}+b_{2}b_{3}+2b_{1}b_{4}-4b_{2}b_{4}\protect\\
 & = & -3b_{1}\bar{b}_{2}-\bar{b}_{2}b_{3}-2b_{1}\bar{b}_{4}-\bar{b}_{2}\bar{b}_{4}+5b_{1}+b_{3}+4\bar{b}_{2}+4\bar{b}_{4}-4.
\end{eqnarray}
The first expression is highly non-submodular while the second is entirely submodular.


\refsec
\begin{itemize}
\item 2018, original paper (Theorem 8): \cite{Boros2018boundsPaper}.
\end{itemize}

\newpage

\subsection{PFR-BCR-1 (Boros, Crama,  Rodr\'{i}guez-Heck, 2018)}

\summarysec

The parity function for even $n$ can be quadratized with:

\begin{align}
f\left(b_{1},b_{2},\ldots,b_{n}\right)\rightarrow1+2\sum_{ij}b_{i}b_{j}-\sum_{i}b_{i}+4\sum_{2i}^{n-1}b_{a_{i}}\left(i-\sum_{j}^{n}b_{j}\right)
\end{align}

\costsec
\begin{itemize}
\item $\lceil \log(n)\rceil -1$ auxiliary variables.
\item $\lfloor 0.5n\rfloor$ non-submodular linear terms. 
\item $ n^2$ non-submodular quadratic terms. 
\end{itemize}

\prossec
\begin{itemize}
\item Smaller number of auxiliary variables than the most naive methods. 
\item Fewer non-submodular linear terms than in the analogous quadratization for its complement (the parity function). 
\end{itemize}

\conssec
\begin{itemize}
\item Only works for a special class of functions
\item Introduces many linear and even more quadratic non-submodular terms (everything is non-submodular except for $0.5n^2$ quadratic terms involving the auxiliaries with the non-auxiliaries, and all $n$ linear terms involving only the non-auxiliaries).
\item non-submodular terms can be rather large compared to the submodular terms (about $4n$ times as big).
\end{itemize}

\examplesec

By bit-flipping $b_2$ and $b_4$, i.e. substituting $b_2 = 1 - \bar{b}_{2}$ and $b_4 = 1 - \bar{b}_{4}$, we see that:
\begin{eqnarray}
H & = & 3b_{1}b_{2}+b_{2}b_{3}+2b_{1}b_{4}-4b_{2}b_{4}\protect\\
 & = & -3b_{1}\bar{b}_{2}-\bar{b}_{2}b_{3}-2b_{1}\bar{b}_{4}-\bar{b}_{2}\bar{b}_{4}+5b_{1}+b_{3}+4\bar{b}_{2}+4\bar{b}_{4}-4.
\end{eqnarray}
The first expression is highly non-submodular while the second is entirely submodular.


\refsec
\begin{itemize}
\item 2018, original paper (Theorem 11): \cite{Boros2018boundsPaper}.
\end{itemize}

\newpage
\subsection{PFR-BCR-2 (Boros, Crama,  Rodr\'{i}guez-Heck, 2018)}

\summarysec

The parity function for odd $n$ can be quadratized with:

\begin{align}
f\left(b_{1},b_{2},\ldots,b_{n}\right)\rightarrow1+2\sum_{ij}b_{i}b_{j}-\sum_{i}b_{i}+4\sum_{2i}^{n-1}b_{a_{i}}\left(i-\sum_{j}^{n}b_{j}\right)
\end{align}

\costsec
\begin{itemize}
\item $\lceil \log(n)\rceil -1$ auxiliary variables.
\item $\lfloor 0.5n\rfloor$ non-submodular linear terms. 
\item $ n^2$ non-submodular quadratic terms. 
\end{itemize}

\prossec
\begin{itemize}
\item Smaller number of auxiliary variables than the most naive methods. 
\item Fewer non-submodular linear terms than in the analogous quadratization for its complement (the parity function). 
\end{itemize}

\conssec
\begin{itemize}
\item Only works for a special class of functions
\item Introduces many linear and even more quadratic non-submodular terms (everything is non-submodular except for $0.5n^2$ quadratic terms involving the auxiliaries with the non-auxiliaries, and all $n$ linear terms involving only the non-auxiliaries).
\item non-submodular terms can be rather large compared to the submodular terms (about $4n$ times as big).
\end{itemize}

\examplesec

By bit-flipping $b_2$ and $b_4$, i.e. substituting $b_2 = 1 - \bar{b}_{2}$ and $b_4 = 1 - \bar{b}_{4}$, we see that:
\begin{eqnarray}
H & = & 3b_{1}b_{2}+b_{2}b_{3}+2b_{1}b_{4}-4b_{2}b_{4}\protect\\
 & = & -3b_{1}\bar{b}_{2}-\bar{b}_{2}b_{3}-2b_{1}\bar{b}_{4}-\bar{b}_{2}\bar{b}_{4}+5b_{1}+b_{3}+4\bar{b}_{2}+4\bar{b}_{4}-4.
\end{eqnarray}
The first expression is highly non-submodular while the second is entirely submodular.


\refsec
\begin{itemize}
\item 2018, original paper (Theorem 11): \cite{Boros2018boundsPaper}.
\end{itemize}

\newpage

\subsection{Lower bounds for SFRs (Anthony, Boros, Crama, Gruber, 2014)}


\begin{itemize}
\item There exist symmetric functaons on $n$ variables for which no quadratization can be done without at least $\Omega\left(\sqrt{n}\right)$  auxiliary variables (Theorem 5.3 from \cite{Anthony2014,Anthony2016}.  
\item There exist symmetric functions on $n$ variables for whcih no quadratization linear in the auxiliaries can be done without at least $\Omega\left( \frac{n}{\log_2(n)}\right)$ auxiliary variables (Theorem 5.5 from \cite{Anthony2014,Anthony2016}).
\item The parity function on $n$ variables cannot be quadratized without quadratic terms involving the auxiliary variables, unless there is at least $\sqrt{\nicefrac{n}{4}-1}+1=\Omega\left( \sqrt{n}  \right)$ auxiliary variables (Theorem 5.6 from \cite{Anthony2014,Anthony2016}).
\item Theorem 5 of \cite{Boros2018boundsPaper} gives an even tighter bound of $\lceil \log(n)\rceil -1$ for the minimum number of auxiliary variables for the parity function.
\item Corollary 5 of \cite{Boros2018boundsPaper} gives $m\ge \log\left( 1/2 - \mu\right) + \log(n) -1$.
\end{itemize}

\subsection{Lower bounds for positive monomials (Boros, Crama, Rodr\'{i}guez-Heck, 2018)}

\begin{itemize}
\item A positive monomial with $n$ variables cannot be quadratized with fewer than $\lceil \log(n)\rceil - 1$ auxiliary variables, unless there is some extra deduction we can make about the optimization problem, as in for example deduc-reduc (Corollary 1 from \cite{Boros2018boundsPaper}).
\item ALCN (at least $c$ out of $n$) and ECN (exact $c$ out of $n$) functions also cannot be quadratized with fewer than $\lceil \log(n)\rceil - 1$ auxiliary variables (Corollaries 2 and 3 from \cite{Boros2018boundsPaper}).
\item ECN (exact $c$ out of $n$) functions also cannot be quadratized with fewer than $\max\left(\lceil \log(c)\rceil , \lceil \log(n-c)\rceil  \right)-1$ auxiliary variables (Corollaries 2 and 3 from \cite{Boros2018boundsPaper}).
\end{itemize}

\newpage

\subsection{Lower bounds for ZUCs (Boros, Crama, Rodr\'{i}guez-Heck, 2018)}

\begin{itemize}
\item There exist ZUC (zero until $c$) functions such that every quadratization must involve at least $\Omega\left(2^{n/2}\right)$ auxiliary variables, no matter what the value of $c$ (Theorem 2 from \cite{Boros2018boundsPaper}). This is true for almost all ZUC functions because the set of ZUCs requiring fewer auxiliary variables has Lebesgue measure zero.
\item For any $c\ge 0$, the number of auxiliary variables is $m\ge \lceil \log(c) \rceil -1$ (Theorem 3 of \cite{Boros2018boundsPaper}).
\end{itemize}

\subsection{Lower bounds for $d$-sublinear functions (Boros, Crama, Rodr\'{i}guez-Heck, 2018)}

\begin{itemize}
\item The number of auxiliary variables $m$ is such that $2^{m+1} \ge \frac{\beta(q_1)}{2} -d +1$ (Theorem 12 from \cite{Boros2018boundsPaper}).
\item For any $c\ge 0$, the number of auxiliary variables is $m\ge \lceil \log(c) \rceil -1$ (Theorem 3 of \cite{Boros2018boundsPaper}).
\end{itemize}

\newpage


\section{Methods that quadratize MULTIPLE terms with the SAME auxiliaries (Case 2: Arbitrary Functions)}
\subsection{Reduction by Substitution (Rosenberg 1975)}
\summarysec

Pick a variable pair $(b_{i},b_{j})$ and substitute $b_{i}b_{j}$ with a  new auxiliary variable $b_{a_{ij}}$.
Enforce equality in the ground states by adding some scalar multiple of the penalty $P=b_{i}b_{j}-2b_{i}b_{a_{ij}}-2b_{j}b_{a_{ij}}+3b_{a_{ij}}$ or similar. Since $P > 0$ if and only if $b_{a_{ij}}\ne b_ib_j$, the minimum of the new $(k-1)$-local function will satisfy $b_{a_ij}=b_{i}b_{j})$, which means that at the minimum, we have precisely the original function. Repeat $(k-2)$ times for each $k$-local term and the resulting function will be 2-local. For an arbitrary cubic term we have:

\begin{equation}
b_ib_j b_k \rightarrow b_ab_k + b_ib_j - 2b_ib_a - 2b_jb_a + 3b_a.
\end{equation}

\costsec
\begin{itemize}
\item 1 auxiliary variable per reduction.
\item  At most $k\,t$  auxiliary variables for a $k$-local objective function of $t$ terms, but usually fewer.
\end{itemize}

\prossec
\begin{itemize}
\item Variable can be used across the entire objective function, reducing many terms at once.
\item Very easy to implement.
\item Reproduces not only the ground state, but the full spectrum.
\end{itemize}

\conssec
\begin{itemize}
\item Inefficient for single terms as it introduces many auxiliary variables compared to Ishikawa reduction, for example.
\item Introduces quadratic terms with large positive coefficients, making them highly non-submodular.
\item Determining optimal substitutions can be expensive.
\end{itemize}

\examplesec

We pick the pair $(b_{1},b_{2})$ and combine.

\begin{equation}
b_{1}b_{2}b_{3}+b_{1}b_{2}b_{4}\mapsto b_{3}b_a+b_{4}b_a+b_{1}b_{2}-2b_{1}b_a-2b_{1}b_a+3b_a
\end{equation}

\refsec
\begin{itemize}
\item Original paper: \cite{Rosenberg1975}
\item Re-discovered in the context of diagonal quantum Hamiltonians: \cite{Biamonte2008a}.
\item Used in: \cite{Perdomo2008, Bian2013}. 
\end{itemize}

\newpage

\subsection{FGBZ Reduction for Negative Terms (Fix-Gruber-Boros-Zabih, 2011)}

\summarysec

We consider a set $C$ of variables which can occur in multiple terms throughout the objective function.
Each application `rips out' this common component from each term \cite{Fix2011,Boros2014}.

\begin{equation}
\sum_{H}\alpha_{H}\prod_{j\in H}b_{j}\rightarrow\sum_{H}\alpha_{H}\left(1-\prod_{j\in C}b_{j}-\prod_{j\in H\setminus C}b_{j}\right)b_a
\end{equation}

\costsec
\begin{itemize}

\item One auxiliary variable per application.
\item In combination with \ref{subsec:Negative-Monomial-Reduction}, it can reduce $t$ positive terms of degree $k$ in $n$ variables using $n+t(k-1)$ auxiliary variables in the worst case.
\end{itemize}

\prossec
\begin{itemize}
\item Can reduce the connectivity of an objective function, as it breaks interactions between variables.
\end{itemize}

\conssec
\begin{itemize}
\item Cannot reduce the degree of the original function if $|C|\le 1$, and cannot quadratize anything for the other values of $|C|$ (but it can reduce their degree).
\end{itemize}

\examplesec

First let $C=b_{1}$ and use the positive weight version:
\begin{eqnarray}
b_{1}b_{2}b_{3}+b_{1}b_{2}b_{4} & \mapsto & 2b_{a_1}b_{1}+(1-b_{a_1})b_{2}b_{3}+(1-b_{a_1})b_{2}b_{4}\\
 & = & 2b_{a_1}b_{1}+b_{2}b_{3}+b_{2}b_{4}-b_{a_1}b_{2}b_{3}-b_{a_1}b_{2}b_{4}
\end{eqnarray}
now we can use \ref{subsec:Negative-Monomial-Reduction}:

\begin{eqnarray}
-b_{a_1}b_{2}b_{3}-b_{a_1}b_{2}b_{4} & \mapsto & 2b_{a_2}-b_{a_1}b_{a_2}-b_{a_2}b_{2}-b_{a_2}b_{3}+2b_{a_2}-b_{a_1}b_{a_2}-b_{a_2}b_{2}-b_{a_2}b_{4}\\
 & = & 4b_{a_2}-2b_{a_1}b_{a_2}-2b_{a_2}b_{2}-b_{a_2}b_{3}-b_{a_2}b_{4}.
\end{eqnarray}

\refsec
\begin{itemize}
\item Original paper and application to image denoising: \citep{Fix2011}.
\end{itemize}

\newpage

\subsection{FGBZ Reduction for Positive Terms (Fix-Gruber-Boros-Zabih, 2011)}

\summarysec

We consider a set $C$ of variables which can occur in multiple terms throughout the objective function.
Each application `rips out' this common component from each term \cite{Fix2011,Boros2014}:



\begin{equation}
\sum_{H}\alpha_H\prod_{j\in H} b_{j} \rightarrow \sum_{H}\alpha_{H}b_a\prod_{j\in C}b_{j}+\sum_{H}\alpha_{H}(1-b_a)\prod_{j\in H\setminus C}b_{j}.
\end{equation}

\costsec
\begin{itemize}

\item One auxiliary variable per application.
\item In combination with \ref{subsec:Negative-Monomial-Reduction}, it can reduce $t$ positive terms of degree $k$ in $n$ variables using $n+t(k-1)$ auxiliary variables in the worst case.
\end{itemize}

\prossec
\begin{itemize}
\item It is a `perfect' transformation, meaning that after minimizing over $b_a$, the original degree-$k$ function is recovered.
\item Can reduce the connectivity of an objective function, as it breaks interactions between variables.
\end{itemize}

\conssec
\begin{itemize}
\item  If $|C|=1$ the first sum will result in a quadratic but the second sum will have degree $k$. If $|C|=k-1$ the second sum will be quadratic but the first term will have degree $k$. For any other value of $|C|$, both sums will be super-quadratic, but the part of the second sum involving $b_a$ will be negative and therefore can be quadratized easily.
\end{itemize}

\examplesec

With $C=b_{1}$ we can get:
\begin{eqnarray}
b_{1}b_{2}b_{3}+b_{1}b_{2}b_{4} & \rightarrow & 2b_{a_1}b_{1}+(1-b_{a_1})b_{2}b_{3}+(1-b_{a_1})b_{2}b_{4}\\
 & = & 2b_{a_1}b_{1}+b_{2}b_{3}+b_{2}b_{4}-b_{a_1}b_{2}b_{3}-b_{a_1}b_{2}b_{4}
\end{eqnarray}
now we can use \ref{subsec:Negative-Monomial-Reduction} to quadratize the two negative cubic terms:

\begin{eqnarray}
-b_{a_1}b_{2}b_{3}-b_{a_1}b_{2}b_{4} & \mapsto & 2b_{a_2}-b_{a_1}b_{a_2}-b_{a_2}b_{2}-b_{a_2}b_{3}+2b_{a_2}-b_{a_1}b_{a_2}-b_{a_2}b_{2}-b_{a_2}b_{4}\\
 & = & 4b_{a_2}-2b_{a_1}b_{a_2}-2b_{a_2}b_{2}-b_{a_2}b_{3}-b_{a_2}b_{4}.
\end{eqnarray}

\refsec
\begin{itemize}
\item Original paper and application to image denoising: \citep{Fix2011}.
\end{itemize}

\newpage

\subsection{Pairwise Covers (Anthony-Boros-Crama-Gruber, 2017)}

\summarysec

Here we consider a set $C$ of variables which occur in multiple monomials throughout the objective function.
Each application 'rips out' this common component from each term \cite{Fix2011}\cite{Boros2014}.

Let $\mathcal{H}$ be a set of monomials, where $C \subseteq H$ for each $H\in\mathcal{H}$ and each monomial $H$ has a weight $\alpha_{H}$.
The algorithm comes in 2 parts: when all $\alpha_{H}>0$ and when all $\alpha_{H}<0$. Combining the 2 gives the final method:

\begin{enumerate}
\item $\alpha_{H}>0$ 
\begin{equation}
\sum_{H\in\mathcal{H}}\alpha_{H}\prod_{j\in H}b_{j}=\min_{b_a}\left(\sum_{H\in\mathcal{H}}\alpha_{H}\right)b_a\prod_{j\in C}b_{j}+\sum_{H\in\mathcal{H}}\alpha_{H}(1-b_a)\prod_{j\in H\setminus C}b_{j}
\end{equation}
\item $\alpha_{H}<0$
\begin{equation}
\sum_{H\in\mathcal{H}}\alpha_{H}\prod_{j\in H}b_{j}=\min_{b_a}\sum_{H\in\mathcal{H}}\alpha_{H}\left(1-\prod_{j\in C}b_{j}-\prod_{j\in H\setminus C}b_{j}\right)b_a
\end{equation}
\end{enumerate}

\costsec
\begin{itemize}

\item One auxiliary variable per application.
\item In combination with \ref{subsec:Negative-Monomial-Reduction}, it can be used to make an algorithm which can reduce $t$ positive monomials of degree $d$ in $n$ variables using $n+t(d-1)$ auxiliary variables in the worst case.
\end{itemize}

\prossec
\begin{itemize}
\item Can reduce the connectivity of an objective function, as it breaks interactions between variables.
\end{itemize}

\conssec
\begin{itemize}
\item $\alpha_{H}>0$ method converts positive terms into negative ones of same order rather than reducing them, though these can then be reduced more easily.
\item $\alpha_{H}<0$ method only works for $|C|>1$, and cannot quadratize cubic terms.
\end{itemize}

\examplesec

First let $C=b_{1}$ and use the positive weight version:
\begin{eqnarray}
b_{1}b_{2}b_{3}+b_{1}b_{2}b_{4} & \mapsto & 2b_{a_1}b_{1}+(1-b_{a_1})b_{2}b_{3}+(1-b_{a_1})b_{2}b_{4}\\
 & = & 2b_{a_1}b_{1}+b_{2}b_{3}+b_{2}b_{4}-b_{a_1}b_{2}b_{3}-b_{a_1}b_{2}b_{4}
\end{eqnarray}
now we can use \ref{subsec:Negative-Monomial-Reduction}:

\begin{eqnarray}
-b_{a_1}b_{2}b_{3}-b_{a_1}b_{2}b_{4} & \mapsto & 2b_{a_2}-b_{a_1}b_{a_2}-b_{a_2}b_{2}-b_{a_2}b_{3}+2b_{a_2}-b_{a_1}b_{a_2}-b_{a_2}b_{2}-b_{a_2}b_{4}\\
 & = & 4b_{a_2}-2b_{a_1}b_{a_2}-2b_{a_2}b_{2}-b_{a_2}b_{3}-b_{a_2}b_{4}.
\end{eqnarray}

\refsec
\begin{itemize}
\item Theorem 4 of: \citep{Anthony2017}.
\end{itemize}

\subsection{Flag Based SAT Mapping\label{sub:flag_SAT}}

\summarysec

This method uses gadgets to produce separate 3-SAT clauses which allow variables which `flag' the state of pairs of other variables.

\costsec
\begin{itemize}
\item 24 auxiliary variables to quadratize $z_1z_1z_3$.
\end{itemize}

\prossec
\begin{itemize}
\item  Very general and therefore conducive to proofs.
\end{itemize}

\conssec
\begin{itemize}
\item Extremely inefficient in terms of number of auxiliary variables.
\end{itemize}

\examplesec

%

To create a system which maps $b_1b_2b_3$,  we use the following gadget (note that this is given in terms of $z$ in the orginal work and translated to $b$ here):
\begin{align}
H_1(b_1,b_2,b_3)=2\sum_{i=1}^3b_ib_{a_i}+2\sum_{i<j}^3b_{a_i}b_{a_j}-4\sum_{i=1}^3b_{a_i}-2\sum_{i=1}^3b_i+\frac{23}{2}.
\end{align}

Implementing $\alpha H_1$, creates a situation where $b_3$ is a `flag' for $b_1$ and $ b_2$ in other words $b_3$ is constrained to be $1$ in the low energy manifold if $b_1=0$ and $ b_2=0$. It follows from the universality of $3-\rm{SAT}$ that these `flag' clauses can be combined to map any spin Hamiltonian. To do this, we also need anti-ferromagnetic couplings to express the `negated' variable, to do this, we define,

\begin{align}
H_2(b_1,b_{\neg1})=2\,b_1b_{\neg1}-b_i-b_{\neg 1}+1.
\end{align}

 As an explicit example, consider reproducing the spectrum of $z_1z_2z_3=(2\,b_1-1)(2\,b_2-1)(2\,b_3-1)$. In this case we need to assign a higher energy to the $(1,1,1)$, $(0,0,1)$, $(0,1,0)$, and  $(1,0,0)$ states. A flag ($b_{a_{4,1}}$) which is forced into a higher energy state if these conditions are satisfied can be constructed from two instances of $H_1$ and an auxilliary qubit, combining these leads to

{\scriptsize
\begin{align}
z_1z_2z_3 \rightarrow & \alpha\left(\sum_{i=1}^3 H_2(b_i,b_{\neg i})+\sum_{i=1}^{2^3} H_2(b_{a_i},b_{\neg a_i})+ H_2(b_{a_{4,1}},b_{a_{4,2}})+H_1(b_{a_{1,1}},b_{a_{1,2}},b_{\neg a_1})+H_1(b_{a_1},b_{a_{1,3}},b_{a_{4,1}}) ~+\right.  \\
 & H_1(b_{1},b_{2},b_{\neg a_2})+H_1(b_{a_2},b_{3},b_{a_{4,2}}+H_1(b_{1},b_{2},b_{\neg a_3})+H_1(b_{a_3},b_{a_{1,3}},b_{a_{4,1}})+H_1(b_{a_{1,1}},b_{a_{1,2}},b_{\neg a_4})~+ \\
 &H_1(b_{a_4},b_{3},b_{a_{4,2}}) +  H_1(b_{1},b_{a_{1,2}},b_{\neg a_5})+H_1(b_{a_5},b_{3},b_{a_{4,1}})+H_1(b_{a_{1,1}},b_{2},b_{\neg a_6})+H_1(b_{a_6},b_{a_{1,3}},b_{a_{4,2}})~+ \\
 & \left.H_1(b_{1},b_{2},b_{\neg a_7})+H_1(b_{a_7},b_{a_{1,3}},b_{a_{4,1}})+H_1(b_{a_{1,1}},b_{ a_{1,2}},b_{\neg a_8})+H_1(b_{a_8},b_{3},b_{a_{4,2}})\right) 
-2\,b_{a_{4,1}}+1.
\end{align}
}
Each of the next four lines assigns a value to the flag variable $b_{a_{4,1}}$ for a state and $b_{a_{4,2}}$, for instance the leftmost two terms of the second line enforce that $b_{a_{4,1}}=0$ if $(b_1,b_2,b_3)=(0,0,0)$, while the right two terms enforce that $b_{a_{4,1}}=1$ if $(b_1,b_2,b_3)=(1,1,1)$. Because there are $2^3=8$ possible bitstrings for $(b_1,b_2,b_3)$, and each term to enforce a flag state requires two instances of $H_1$ (and two auxilliary variables), a total of $16$ instances are required as well as $16$ auxilliary variables.

\refsec
\begin{itemize}
\item Paper showing the universality of the Ising spin models: \cite{DelasCuevasGemmaandCubitt2016}.
\end{itemize}

\newpage

\subsection{Lower bounds for arbitrary functions (Anthony, Boros, Crama, Gruber, 2015)}

\begin{itemize}
\item There exist functions on $n$ variables for which no quadratization can be done without at least $\frac{2^{\nicefrac{n}{2}}}{8}=\Omega\left(\sqrt{n}\right)$  auxiliary variables (Theorem 5.3 from \cite{Anthony2017}.  
\item There exist symmetric functions on $n$ variables for which no quadratization linear in the auxiliaries can be done without at least $\Omega\left( \frac{2^n}{n}\right)$ auxiliary variables (Theorem 5.5 from \cite{Anthony2017}).
\end{itemize}

\newpage

\section{Strategies for combining methods}

\subsection{SCM-BCR (Boros, Crama, and Rodr\'{i}guez-Heck, 2018)}

\summarysec

Split a $k$-local monomial with odd $k$ into a $(k-1)$-local term (with even degree) and a new odd $k$-local term which has negative coefficient:

\begin{align}
\begin{gathered}
b_{1}b_{2}\cdots b_{k}	\rightarrow \prod_{i=1}^{k-1}b_{i}-\prod_{i=1}^{k-1}b_{i}(1-b_{k})
\end{gathered}
\end{align}

\noindent We can use any of the PTR methods for even $k$ on the first term, and we can use any of the NTR methods on the second term. Can be generalized to split into different-degree factors when seeking an "optimum" quadratization. Can be generalized into more splits.

\costsec
\begin{itemize}
\item Depends on the methods used for the PTR and NTR procedures.
\end{itemize}

\prossec
\begin{itemize}
\item Very flexible.
\end{itemize}

\conssec
\begin{itemize}
\item First turns one term into two terms, so might not be preferred when we wish to minimize the number of terms.
\end{itemize}

\refsec
\begin{itemize}
\item Original paper: \cite{Boros2018QuadratizationsOS}.
\end{itemize}

\newpage

\subsection{Decomposition into symmetric and anti-symmetric parts}

\summarysec

Split a any function $f$ into a symmetric part and anti-symmetric part:

\begin{align}
f\left(b_1,b_2,\ldots,b_n\right) &=  f_{\textrm{symmetric}} + f_{\textrm{anti-symmetric}}  ,\\
f_{\textrm{symmetric}}                                 &\equiv \frac{1}{2}\left( f\left(b_1,b_2,\ldots,b_n\right) + f\left(1-b_1,1-b_2,\ldots,1-b_n\right)  \right) \\
f_{\textrm{anti-symmetric}} &\equiv \frac{1}{2}\left( f\left(b_1,b_2,\ldots,b_n\right) - f\left(1-b_1,1-b_2,\ldots,1-b_n\right)  \right) 
\end{align}

\noindent  We can now use any of the methods described only for symmetric functions, on the symmetric part, and use the (perhaps less powerful) general methods on the anti-symmetric part.

\costsec
\begin{itemize}
\item Depends on the methods used.
\end{itemize}

\prossec
\begin{itemize}
\item Allows non-symmetric functions to benefit from techniques designed only for symmetric functions.
\end{itemize}

\conssec
\begin{itemize}
\item May result in more terms than simply quadratizing the non-symmetric function directly.
\end{itemize}

\refsec
\begin{itemize}
\item Discussed in: \cite{Kahl2011}.
\end{itemize}

\newpage

%

\newpage

\part{\underline{{\normalsize Hamiltonians quadratic in $z$ and linear in $x$ (Transverse Field Ising Hamiltonians)}}\label{partTransverseIsing}}

The Ising Hamiltonian with a transverse field in the $x$ direction is possible to implement in hardware:

\begin{align}
H = \sum_i \left( \alpha_i^{(z)} z_i + \alpha_i^{(x)} x_i\right) + \sum_{ij}\left(\alpha_{ij}^{(zz)} z_iz_j  \right).
\end{align}	

\subsection{ZZZ-TI-CBBK: Transvese Ising from ZZZ, by Cao, Babbush, Biamonte, and Kais (2015)}

There is only one reduction in the literature for reducing a Hamiltonian term to the transverse Ising Hamiltonian, and it works on 3-local $zzz$ terms, by introducing an auxiliary qubit with label $a$:

\begin{align}
\alpha z_{i}z_{j}z_{k}\rightarrow\alpha^{I}+\alpha_{i}^{z}z_{i}+\alpha_{j}^{z}z_{j}+\alpha_{k}^{z}z_{k}+\alpha_{a}^{z}z_{a}+\alpha_{a}^{x}x_{a}+\alpha_{ia}^{zz}z_{i}z_{a}+\alpha_{ja}^{zz}z_{j}z_{a}+\alpha_{ka}^{zz}z_{k}z_{a} \label{eq:ZZZ-TI-CBBK} 
\end{align}

\begin{tabular}{rcl}
$\alpha^{I}$ & = & $\frac{1}{2}\left(\Delta\textcolor{red}{\ensuremath{+}}\left(\frac{\alpha}{6}\right)^{\nicefrac{2}{5}}\Delta^{\nicefrac{3}{5}}\right)$\tabularnewline
$\alpha_{i}^{z}$ & = & $-\frac{1}{2}\left(\left(\frac{7\alpha}{6}+\left(\frac{\alpha}{6}\right)^{\nicefrac{3}{5}}\Delta^{\nicefrac{2}{5}}\right)\textcolor{red}{\ensuremath{-}}\left(\frac{\alpha\Delta^{4}}{6}\right)^{\nicefrac{1}{5}}\right)$\tabularnewline
$\alpha_{j}^{z}$ & = & $\alpha_{i}^{(z)}$\tabularnewline
$\alpha_{k}^{z}$ & = & $\alpha_{i}^{(z)}$\tabularnewline
$\alpha_{a}^{z}$ & = & $\frac{1}{2}\left(\Delta\textcolor{red}{\ensuremath{-}}\left(\frac{\alpha}{6}\right)^{\nicefrac{2}{5}}\Delta^{\nicefrac{3}{5}}\right)$\tabularnewline
$\alpha_{a}^{x}$ & = & $\left(\frac{\alpha\Delta^{4}}{6}\right)^{\nicefrac{1}{5}}$\tabularnewline
$\alpha_{ia}^{zz}$ & = & $-\frac{1}{2}\left(\left(\frac{7\alpha}{6}+\left(\frac{\alpha}{6}\right)^{\nicefrac{3}{5}}\Delta^{\nicefrac{2}{5}}\right)\textcolor{red}{\ensuremath{+}}\left(\frac{\alpha\Delta^{4}}{6}\right)^{\nicefrac{1}{5}}\right)$\tabularnewline
$\alpha_{ja}^{zz}$ & = & $\alpha_{ia}^{(zz)}$\tabularnewline
$\alpha_{ka}^{zz}$ & = & $\alpha_{ja}^{(zz)}$\tabularnewline
\end{tabular}

\vspace{5mm}
Including all coefficients and factorizing, we get:

\begin{align}
\alpha z_iz_jz_k \rightarrow & \left( \Delta + \frac{\alpha \Delta^4}{6}^{\nicefrac{1}{5}} \left(z_i+z_j+z_k \right)  \right)\left( \frac{1-z_a}{2}\right) + \frac{\alpha \Delta^4}{6}^{\nicefrac{1}{5}}  x_a\\
&+ \left( \left(\frac{\alpha}{6}\right)^{\nicefrac{2}{5}} \Delta^{\nicefrac{3}{5}} - \left( \frac{7\alpha}{6} + \left(\frac{\alpha}{6} \right)^{\nicefrac{3}{5}} \Delta^{\nicefrac{2}{5}} \right)\left(z_i + z_j + z_k \right)        \right)\left(\frac{1+z_a}{2} \right)
\end{align}

The low-lying spectrum (eigenvalues \textbf{\textit{and}} eigenvectors) of the right side of Eq. \eqref{eq:ZZZ-TI-CBBK} will match those of the left side to within a spectral error of $\epsilon$ as long as $\Delta = \mathcal{O}\left(\epsilon^{-5}\right)$.

\costsec
\begin{itemize}
\item 1 auxiliary qubit
\item 8 auxiliary terms not proportional to \openone.
\end{itemize}


%
%
%
%
%
%

\newpage

\part{\underline{{\normalsize{General Quantum Hamiltonians}}}}\label{partGeneral}

\section{Non-perturbative Gadgets}

\subsection{NP-OY (Ocko \& Yoshida, 2011)}

\summarysec

For the 8-body Hamiltonian:

\begin{eqnarray}
\begin{gathered}
H_{8\textrm{-body}}	=-J\sum_{ij}\left(x_{ij3}x_{ij+1,2}x_{ij4}x_{ij+1,4}x_{i+1j1}x_{i+1j+1,1}x_{i+1j3}x_{ij+1,2}+\right. \\
	\left.z_{ij1}z_{ij2}z_{ij3}z_{ij4}+z_{i-1j4}z_{ij1}+z_{ij2}z_{ij-1,3}+z_{ij4}z_{i+1j1}+z_{ij3}z_{ij+1,2}\right),
\end{gathered}
\end{eqnarray}

\noindent we define auxiliary qubits labeled by $a_{ijk}$, two auxiliaries for each pair $ij$: labeled $a_{ij1}$ and $a_{ij2}$. Then the 8-body Hamiltonian has the same low-lying eigenspace as the 4-body Hamiltonian:

\begin{eqnarray}
\begin{gathered}
H_{4\textrm{-body}}	=-\sum_{ij}\alpha\left(z_{ij1}z_{ij2}z_{ij3}z_{ij4}+z_{i,j,-1,4}z_{ij1}+z_{ij2}z_{i,j-1,3}+z_{ij4}z_{i+1,j,1}+z_{ij3}z_{i,j+1,2}\right.\\
	\left(1-z_{a_{ij1}}+z_{a_{ij2}}+z_{a_{ij1}}z_{a_{ij2}}\right)\left(z_{a_{i,j+1,1}}+z_{a_{i,j+1,2}}+z_{a_{i,j+1,1}}z_{a_{i,j+1,2}}-1\right)+\\
	\left.\left(1+z_{a_{ij1}}-z_{a_{ij2}}+z_{a_{ij1}}z_{a_{ij2}}\right)\left(1-z_{a_{i+1,j1}}-z_{a_{i+1,j2}}-z_{a_{i+1,j1}}z_{a_{i+1,j2}}\right)\right)+\\
	\frac{U}{2}\left(z_{a_{ij1}}+z_{a_{ij2}}+z_{a_{ij1}}z_{a_{ij2}}-1\right)+\\
	\frac{t}{2}\left(\left(x_{a_{ij2}}+z_{a_{ij1}}x_{a_{ij2}}\right)x_{ij3}x_{ij4}+\left(x_{a_{ij1}}x_{a_{ij2}}+y_{a_{ij1}}y_{a_{ij2}}\right)x_{i,j+1,2}x_{i,j+1,4}+\right.\\
\left.	\left.\left(x_{a_{ij2}}-z_{a_{ij1}}x_{a_{ij2}}\right)x_{i+1,j+1,1}x_{i+1,j+1,2}+\left(x_{a_{ij1}}x_{a_{ij2}}-y_{a_{ij1}}y_{a_{ij2}}\right)x_{i+1,j,1}x_{i+1,j,3}\right)\right).
\end{gathered}
\end{eqnarray}

\noindent Now by defining the following ququits (spin-${3/2}$ particles, or 4-level systems):

\begin{align}
s_{ijki^{\prime}j^{\prime}k^{\prime}}^{zz}	&=z_{ijk}z_{i^{\prime}j^{\prime}k^{\prime}}\\
s_{a_{ij}1}^{zz}	&=\left(1-z_{a1_{ij}}+z_{a2_{ij}}+z_{a1_{ij}}z_{a2_{ij}}\right)\\
s_{a_{ij}2}^{zz}	&=\left(z_{a1_{ij}}+z_{a2_{ij}}+z_{a1_{ij}}z_{a2_{ij}}-1\right)\\
s_{a_{ij}3}^{zz}	&=\left(1+z_{a1_{ij}}-z_{a2_{ij}}+z_{a1_{ij}}z_{a2_{ij}}\right)\\
s_{a_{ij}1}^{xz}	&=\left(x_{a2_{ij}}+z_{a1_{ij}}x_{a2_{ij}}\right)\\
s_{a_{ij}2}^{xz}	&=\left(x_{a2_{ij}}-z_{a1_{ij}}x_{a2_{ij}}\right)\\
s_{ijki^{\prime}j^{\prime}k^{\prime}}^{xx}	&=x_{ijk}x_{i^{\prime}j^{\prime}k^{\prime}}\\
s_{a_{ij}1}^{xy}	&=\left(x_{a1_{ij}}x_{a2_{ij}}+y_{a1_{ij}}y_{a2_{ij}}\right)\\
s_{a_{ij}2}^{xy}	&=\left(x_{a1_{ij}}x_{a2_{ij}}-y_{a1_{ij}}y_{a2_{ij}}\right)
\end{align}

\subsection*{NP-OY (Ocko \& Yoshida, 2011) [Continued]} 
\noindent We can write the 4-body Hamiltonian on qubits as a 2-body Hamiltonian on ququits:

{\scriptsize
\begin{eqnarray}
\begin{gathered}
H_{2\textrm{-body}}	=-\sum_{ij}\left(\alpha\left(s_{ij1ij2}^{zz}s_{ij3ij4}^{zz}+s_{ij-1,4ij1}^{zz}+s_{ij2ij-1,3}^{zz}+s_{ij4i+1j1}^{zz}+s_{ij3ij+1,2}^{zz}+s_{a_{ij}1}^{zz}s_{a_{ij+1}2}^{zz}-s_{a_{ij}3}^{zz}s_{a_{i+1j}3}^{zz}\right)\right.\\
	\left.+\frac{U}{2}s_{a_{ij},1}^{zz}+\frac{t}{2}\left(s_{a_{ij}1}^{xz}s_{ij3ij4}^{xx}+s_{a_{ij}1}^{xy}s_{ij+1,2ij+1,4}^{xx}+s_{a_{ij}2}^{xz}s_{i+1,j+1,1i+1,j+1,2}^{xx}+s_{a_{ij}2}^{xy}s_{i+1j1,i+1,j3}^{xx}\right)\right).
\end{gathered}
\end{eqnarray}
}

\noindent The low-lying eigenspace of $H_{2-\textrm{body}}$ is \textit{exactly} the same as for $H_{4-\textrm{local}}$.


\costsec
\begin{itemize}
\item 2 auxiliary ququits for each pair $ij$.
\item 6 more total terms (6 terms in the 8-body version becomes 12 terms: \\
11 of them 2-body and 1 of them 1-body).
\end{itemize}

\prossec
\begin{itemize}
\item Non-perturbative. No prohibitive control precision requirement.
\item Only two auxiliaries required for each pair $ij$.
\item 8-body to 2-body transformation can be accomplished in 1 step, rather than a 1B1 gadget which would take 6 steps or an SD + $(3\rightarrow2)$ gadget combination which would take 4 steps.
\end{itemize}
 
\conssec
\begin{itemize}
\item Increase in dimention from working with only 2-level systems (spin-1/2 particles or $2\times2$ matrices) to working with 4-level systems (spin-3/2 particles).
\item Until now, only derived for a very specific Hamiltonian form.
\item This appraoch may become more demanding for Hamiltonians that are more than 8-local.
\end{itemize}


\refsec
\begin{itemize}
\item Original paper: \cite{Ocko2011}.
\end{itemize}

\newpage

\subsection{NP-SJ (Subasi \& Jarzynski, 2016)}



\summarysec

Determine the $k$-local term, $H_{k-\rm{local}}$, whose degree we wish to reduce, and factor it into two commuting factors: $H_{k^\prime-\rm{local}}H_{(k-k^\prime)-\rm{local}}$, where $k^\prime$ can be as low as 0. Separate all terms that are at most $(k-1)$-local into ones that commmute with one of these factors (it does not matter which one, but without loss of generality we assume it to be the $(k-k^\prime)$-local one)  and ones that anti-commute with it:

\begin{align}
 H_{ <k \rm{-local}}^{\rm{commuting}} +H_{ <k \rm{-local}}^{\rm{anti-commuting}}  + \alpha H_{k^\prime-\rm{local}}H_{(k-k^\prime)-\rm{local}}
\end{align}
	
\noindent Introduce one auxiliary qubit labeled by $a$ and the Hamiltonian:

\begin{align}
\alpha x_aH_{k^\prime-\rm{local}} +  H_{ <k \rm{-local}}^{\rm{commuting}} + z_aH_{ <k \rm{-local}}^{\rm{anti-commuting}} 
\end{align}

\noindent no longer contains $H_{k-\rm{local}}$ but $H_{<k-{\rm{local}}}^{\rm{anti-commuting}}$ is now one degree higher.

\costsec
\begin{itemize}
\item 1 auxiliary qubit to reduce $k$-local term to $(k^\prime+1)$-local where $k^\prime$ can even be 0-local, meaning the $k$-local term is reduced to a 1-local one.
\item Raises the $k$-locality of $H_{ <k \rm{-local}}^{\rm{anti-commuting}}$ by 1 during each application. It can become $(>k)$-local!
\end{itemize}

\prossec
\begin{itemize}
\item Non-perturbative
\item Can linearize a term of arbitrary degree in one step.
\item Requires very few auxiliary qubits.
\end{itemize}

\conssec
\begin{itemize}
\item Can introduce many new non-local terms as an expense for reducing only one $k$-local term.
\item If the portion of the Hamiltonian that does not commute with the $(k-k^\prime)$-local term has termms of degree $k-1$ (which can happen if $k^\prime=0$) they will all become $k$-local, so there is no guarantee that this method reduces $k$-locality.
\item If any terms were more than 1-local, this method will not fully quadratize the Hamiltonian (it must be combined with other methods).
\item It only works when the Hamiltonian's terms of degree at most $k-1$  all either commute or anti-commmute with the $k$-local term to be eliminated.
\end{itemize}

\vspace{-1mm}

\examplesec
\vspace{-3mm}

\begin{align}
\vspace{-1mm}
4z_5 -3 x_1 + 2z_1y_2x_5 + 9x_1x_2x_3x_4 -x_1y_2z_3x_5 \rightarrow 9x_{a_{1}} + 4z_{a_2}z_5   -3z_{a_3}x_1  -z_{a_3}x_{a_2}   +2x_{a_3}x_5  
\end{align}

\refsec
\begin{itemize}
\item Original paper, and description of the choices of terms and factors used for the given example \cite{Subas2016}.
\end{itemize}

\subsection{NP-Nagaj-1 (Nagaj, 2010)}

\summarysec
\vspace{-1mm}
The Feynman Hamiltonian can be written as \cite{Feynman1985a}:
\vspace{-5mm}

\begin{align}
\frac{1}{4}\left(x_{1}x_{2}-{\rm i}y_{1}x_{2}+{\rm i}x_{1}y_{2}+y_{1}y_{2}\right)U_{{\rm 2-local}}+\frac{1}{4}\left(x_{1}x_{2}+{\rm i}y_{1}x_{2}-{\rm i}x_{1}y_{2}+y_{1}y_{2}\right)U_{{\rm 2-local}}^{\dagger},
\end{align}

\noindent where $U_{{\rm 2-{\rm local}}}$ is an arbitrary 2-local unitary matrix that acts on qubits different from the ones labeled by "1" and "2". This Hamiltonian that is 4-local on qubits can be transformed into one that is 2-local in qubits and qutrits. Here we show the 2-local Hamiltonian for the case where $U_{{\rm 2-local}}={\rm CNOT}\equiv\frac{1}{2}\left(\openone+z_{3}+x_{4}-z_{3}x_{4}\right).$ We start with the specific 4-local Hamiltonian:

\vspace{-5mm}

\begin{align}
H_{{\rm 4-local}}=\frac{1}{4}\left(x_{1}x_{2}+y_{1}y_{2}+x_{1}x_{2}z_{3}+x_{1}x_{2}x_{4}+y_{1}y_{2}z_{3}+y_{1}y_{2}x_{4}-x_{1}x_{2}z_{3}x_{4}-y_{1}y_{2}z_{3}x_{4}\right),
\end{align}

\noindent and after adding 4 auxiliary qubits labeled by $a_{1}$ to $a_{4}$ and 6 auxiliary qutrits labeled by $a_{5}$ to $a_{10}$ and acted on by the Gell-Mann matrices $\lambda_{1}$ to $\lambda_{9}$, we get the following 2-local Hamiltonian:

\vspace{-5mm}

\begin{align}
H_{{\rm 2-local}}      &=\nicefrac{1}{2}\left(2\lambda_{6,a_{8}} + x_{1}\lambda_{1,a_{5}}+y_{1}\lambda_{2,a_{5}}+\lambda_{6,a_{5}}-z_{3}\lambda_{6,a_{5}}+x_{a_{1}}\lambda_{4,a_{5}}+y_{a_{1}}\lambda_{5,a_{5}}+x_{a_{1}}\lambda_{1,a_{6}} + \right.\\
                       &y_{a_{1}}\lambda_{2,a_{6}}+2x_{4}\lambda_{6,a_{6}}+x_{a_{2}}\lambda_{4,a_{6}}+y_{a_{2}}\lambda_{5,a_{6}}+x_{a_{2}}\lambda_{1,a_{7}}+y_{a_{2}}\lambda_{2,a_{7}}+\lambda_{6,a_{7}}-z_{a_{1}}\lambda_{6,a_{7}}+\\
&x_{2}\lambda_{4,a_{7}}+y_{2}\lambda_{5,a_{7}}+x_{1}\lambda_{1,a_{8}}+ 	y_{1}\lambda_{2,a_{8}}+z_{3}\lambda_{6,a_{8}}+x_{a_{5}}\lambda_{4,a_{9}}+y_{a_{5}}\lambda_{5,a_{9}}+\lambda_{6,a_{9}}+ \\
&\left.x_{a_{6}}\lambda_{4,a_{9}}+y_{a_{6}}\lambda_{5,a_{9}}+x_{a_{6}}\lambda_{1,a_{10}}+y_{a_{6}}\lambda_{2,a_{10}}+\lambda_{6,a_{10}}+z_{3}\lambda_{6,a_{10}}+x_{2}\lambda_{4,a_{10}}+y_{2}\lambda_{5,a_{10}}\right),
\end{align}

\noindent whose low-lying  spectrum is equivalent to the spectrum of  $H_{2-{\rm local}}$.

\costsec
\begin{itemize}
\item 6 auxiliary qutrits and 4 auxiliary qubits
\item 2 quartic, 4 cubic, and 2 quadratic terms becomes 27 quadratic terms and 5 linear terms in the Pauli-GellMann basis. 
\end{itemize}

\prossec
\begin{itemize}
\item Exact (non-perturbative). No special control precision demands.
\item All coefficients are equal to each other, with a value of $\nicefrac{1}{2}$, except one which is equal to 1.
\item With more auxiliary qubits, can be further reduced to only containing qubits.
\end{itemize}

\conssec
\begin{itemize}
\item Involves qutrits in all 32 terms.
\item Only derived (so far) for the Feynman Hamiltonian.
\item High overhead in terms of number of auxiliary qubits and number of terms.
\end{itemize}

\examplesec

The transformation presented above was for the case of $U_{2-\textrm{local}}={\rm CNOT}\equiv\frac{1}{2}\left(\openone+z_{3}+x_{4}-z_{3}x_{4}\right)$, but similar transformations can be derived for any arbitrary unitary matrix $U_{2-\textrm{local}}$.

\refsec
\begin{itemize}
\item Original paper: \cite{Nagaj2010}.
\end{itemize}

\newpage

\subsection{NP-Nagaj-2 (Nagaj, 2012)}

\summarysec

Similar to NP-Nagaj-1 but instead of using qutrits, we use two qubits  for each qutrit, according to:

\vspace{-7mm}

\begin{align}
|0\rangle \rightarrow |00\rangle , \qquad  |1\rangle \rightarrow  \frac{1}{\sqrt{2}}\left( |01\rangle + |10\rangle  \right),  \qquad |2\rangle \rightarrow \frac{1}{\sqrt{2}} \left( |01\rangle - |10\rangle   \right).
\end{align}

\noindent which leads to the following transformations:
\vspace{-5mm}

\begin{align}
|01\rangle\langle10|_{ij}+h.c. &\rightarrow \frac{1}{\sqrt{2}}\left(|01\rangle\langle10|_{ij_{1}}+|01\rangle\langle10|_{ij_{2}}\right)+h.c.\\
|02\rangle\langle10|_{ij}+h.c. &\rightarrow \frac{1}{\sqrt{2}}\left(|01\rangle\langle10|_{ij_{1}}-|01\rangle\langle10|_{ij_{2}}\right)+h.c.\\
|1\rangle\langle2|_{j}+h.c.    &\rightarrow z_{j_{1}}-z_{j_{2}},
\end{align}

\noindent and the following 2-local Hamiltonian involving only qubits:
\vspace{-3mm}

\begin{eqnarray}
\begin{gathered}
H_{2-\textrm{local}}=\nicefrac{1}{2}\left(z_{a_{5}}-z_{a_{6}}+z_{a_{9}}-z_{a_{10}}-z_{a_{3}}z_{a_{5}}+z_{a_{3}}z_{a_{6}}-z_{a_{3}}z_{a_{9}}+z_{a_{3}}z_{a_{10}}+\right. \\
\left.z_{a_{11}}-z_{a_{12}}+z_{a_{15}}-z_{a_{16}}+z_{a_{3}}z_{a_{11}}-z_{a_{3}}z_{a_{12}}+z_{a_{3}}z_{a_{15}}-z_{a_{3}}z_{a_{16}}\right)+x_{4}z_{a_{7}}-x_{4}z_{a_{8}}+z_{a_{13}}-z_{a_{14}}+\\
\nicefrac{1}{\sqrt{2}}\left(x_{1}\lambda_{1,a_{5}}+y_{1}\lambda_{2,a_{5}}+x_{1}\lambda_{1,a_{6}}+y_{1}\lambda_{2,a_{6}}+x_{a_{1}}\lambda_{1,a_{7}}+y_{a_{1}}\lambda_{2,a_{7}}+x_{a_{1}}\lambda_{1,a_{8}}+y_{1}\lambda_{2,a_{8}}\right.+\\
x_{a_{2}}\lambda_{1,a_{7}}+y_{a_{2}}\lambda_{2,a_{7}}+x_{a_{2}}\lambda_{1,a_{8}}+y_{a_{2}}\lambda_{2,a_{8}}+x_{a_{2}}\lambda_{1,a_{9}}+y_{a_{2}}\lambda_{2,a_{9}}+x_{a_{2}}\lambda_{1,a_{10}}+\\
y_{a_{2}}\lambda_{2,a_{10}}+x_{1}\lambda_{1,a_{11}}+y_{1}\lambda_{2,a_{11}}+x_{a_{4}}\lambda_{1,a_{15}}+y_{a_{4}}\lambda_{2,a_{15}}+x_{a_{1}}\lambda_{4,a_{5}}+y_{a_{1}}\lambda_{5,a_{5}}+\\
x_{a_{2}}\lambda_{4,a_{7}}+y_{a_{2}}\lambda_{5,a_{7}}+x_{2}\lambda_{4,a_{9}}+y_{2}\lambda_{5,a_{9}}+x_{a_{3}}\lambda_{4,a_{11}}+y_{a_{3}}\lambda_{5,a_{11}}+x_{a_{4}}\lambda_{4,a_{13}}-\\
y_{a_{4}}\lambda_{5,a_{13}}+x_{a_{2}}\lambda_{4,a_{15}}+y_{a_{2}}\lambda_{5,a_{15}}-x_{a_{1}}\lambda_{4,a_{6}}-y_{a_{1}}\lambda_{5,a_{6}}-x_{a_{2}}\lambda_{4,a_{8}}-y_{a_{2}}\lambda_{5,a_{8}}-x_{2}\lambda_{4,a_{10}}-\\
\left.-y_{2}\lambda_{5,a_{10}}-x_{a_{3}}\lambda_{4,a_{12}}-y_{a_{3}}\lambda_{5,a_{12}}-x_{a_{4}}\lambda_{4,a_{14}}-y_{a_{4}}\lambda_{5,a_{14}}-x_{2}\lambda_{4,a_{16}}-y_{2}\lambda_{5,a_{16}}\right),
\end{gathered}
\end{eqnarray}

\vspace{2mm}

\noindent whose low-lying  spectrum is equivalent to the spectrum of  $H_{2-{\rm local}}$.

\costsec
\begin{itemize}
\item 16 auxiliary qubits.
\end{itemize}

\prossec
\begin{itemize}
\item Exact (non-perturbative). No special control precision demands.
\item Only involves qubits (as opposed to NP-Nagaj-1 which contains qutrits and NP-OY which contains ququits.
\end{itemize}

\conssec
\begin{itemize}
\item Only derived (so far) for the Feynman Hamiltonian.
\item High overhead in terms of number of auxiliary qubits and number of terms.
\end{itemize}

\examplesec

The transformation presented above was for the case of $U_{2-\textrm{local}}={\rm CNOT}\equiv\frac{1}{2}\left(\openone+z_{3}+x_{4}-z_{3}x_{4}\right)$, but similar transformations can be derived for any arbitrary unitary matrix $U_{2-\textrm{local}}$.

\refsec
\begin{itemize}
\item Original paper: \cite{Nagaj2012}.
\end{itemize}

\newpage

\section{Perturbative $(3\rightarrow2)$ Gadgets}

The first gadgets for arbitrary Hamiltonians acting on some number of qubits, were designed to reproduce the spectrum of a 3-local Hamiltonian in the low-lying spectrum of a 2-local Hamiltonian.

\subsection{P$(3\rightarrow2)$-DC  (Duan, Chen, 2011)} 


\summarysec

 For any group of 3-local terms that can be factored into a product of three 1-local factors, we can define three auxiliary qubits (regardless of the number of qubits we have in total) labeled by $a_{i}$  and  make the transformation:


\begin{align}
\prod_i^3 \sum_j \alpha_{ij}s_{i} \rightarrow \alpha + \alpha_i^{ss} \sum_i \left(\sum_j \alpha_{ij}s_{ij}\right)^2 + \alpha_i^{sx}\sum_i \sum_j \alpha_{ij} s_{ij}x_{a_i} + \alpha^{zz} \sum_{ij} z_{a_i}z_{a_j} 
\end{align}

\begin{align}
\alpha    &= \frac{1}{8\Delta} \\
\alpha^{ss} &= \frac{1}{6\Delta^{\nicefrac{1}{3}}} \\
\alpha^{sx} &= - \frac{1}{6\Delta^{\nicefrac{2}{3}}} \\
\alpha^{zz} &= -  \frac{1}{24\Delta}
\end{align}

\noindent The result will be a 2-local Hamiltonian whose low-lying spectrum is equivalent to the spectrum of $H_{3-\rm{local}}$ to within $\epsilon$ as long as $\Delta=\Theta\left(\epsilon^{-3}\right)$. 

\costsec

\begin{itemize}
\item 1 auxiliary qubit for each group of 3-local terms that can be factored into three 1-local factors. 
\item $\Delta =\Theta\left(\epsilon^{-3}\right)$
\end{itemize}

\prossec 
\begin{itemize}
\item Very few auxiliary qubits needed
\end{itemize}

\conssec
\begin{itemize}
\item Will not work for Hamiltonians that do not factorize appropriately.
\end{itemize}  
               
%
%

\refsec

\begin{itemize}
\item Original paper: \cite{Duan2011}
\end{itemize}

\newpage

\subsection{P$(3\rightarrow2)$-DC2 (Duan, Chen, 2011)} 

\summarysec

 For any 3-local term (product of Pauli matrices $s_i$) in the Hamiltonian, we can define \textit{one} auxiliary qubit labeled by $a$  and  make the transformation:                                                                                                                                      


\begin{align}
a\prod_i^3 s_{i} \rightarrow \alpha + \alpha^s s_3 + \alpha^z z_a +\alpha^{ss} \left(s_1 + s_2 \right)^2 + \alpha^{sz}s_3z_a  +  + \alpha^{sx}\left(s_1x_a + s_2x_a  \right) 
\end{align}

\begin{align}
\begin{pmatrix}\alpha & \alpha^{s} & \alpha^{z}\\
\alpha^{ss} & \alpha^{sz} & \alpha^{sx}
\end{pmatrix}=\begin{pmatrix}-\frac{1}{2\Delta} & a\left(\frac{1}{4\Delta^{\nicefrac{2}{3}}}-1\right) & a\left(\frac{1}{4\Delta^{\nicefrac{2}{3}}}-1\right)\\
\frac{1}{\Delta^{\nicefrac{1}{3}}} & \frac{a}{4\Delta^{\nicefrac{2}{3}}} & \frac{1}{\Delta^{\nicefrac{2}{3}}}
\end{pmatrix}
\end{align}

\noindent The result will be a 2-local Hamiltonian whose low-lying spectrum is equivalent to the spectrum of $H_{3-\rm{local}}$ to within $\epsilon$ as long as $\Delta=\Theta\left(\epsilon^{-3}\right)$. 

\costsec

\begin{itemize}
\item 1 auxiliary qubit for each 3-local term. 
\item $\Delta =\Theta\left(\epsilon^{-3}\right)$ 
\end{itemize}

\examplesec


{\small
\begin{align}
x_1z_2y_3 - 3x_1x_2y_4 + z_1x_2 &\rightarrow \alpha +  \alpha^z(z_{a_1} + z_{a_2}) + \alpha^y(y_3 + y_4) +\alpha^{zx}_{12}z_1x_2+   \alpha^{zx}z_2x_{a_1} +\alpha^{xx}_{11} x_1x_2\\
&   +  \alpha^{xx}\left( x_1x_{a_1} + x_1x_{a_2} + x_2x_{a_2} \right)  + \alpha^{yz}\left( y_3z_{a_1} + y_4z_{a_2}  \right)
\end{align}
}

\refsec

\begin{itemize}
\item Original paper: \cite{Duan2011}
\end{itemize}

\newpage

\subsection{P$(3\rightarrow2)$-KKR (Kempe, Kitaev, Regev, 2004)}

\summarysec
 For any 3-local term (product of commuting matrices $s_i$) in the Hamiltonian, we can define three auxiliary qubits labeled by $a_{i}$  and  make the transformation:


\begin{align}
\prod_i^3 s_{i} \rightarrow \alpha   + \alpha_i^{ss} \sum_i s_i^2 + \alpha_i^{sx}\sum_i  s_ix_{a_i} + \alpha^{zz} \sum_{ij} z_{a_i}z_{a_j} 
\end{align}

\begin{align}
\alpha      &= -\frac{1}{8\Delta} \\
\alpha^{ss} &= -\frac{1}{6\Delta^{\nicefrac{1}{3}}} \\
\alpha^{sx} &=  \frac{1}{6\Delta^{\nicefrac{2}{3}}} \\
\alpha^{zz} &=  \frac{1}{24\Delta}
\end{align}

\noindent The result will be a 2-local Hamiltonian whose low-lying spectrum is equivalent to the spectrum of $H_{3-\rm{local}}$ to within $\epsilon$ as long as $\Delta=\Theta\left(\epsilon^{-3}\right)$. 

\costsec

\begin{itemize}
\item 3 auxiliary qubits for each 3-local term. 
\item $\Delta =\Omega\left(\epsilon^{-3}\right)$
\end{itemize}

\examplesec

{\tiny
\begin{align}
x_1z_2y_3 - 3x_1x_2y_4 + z_1x_2 \rightarrow \alpha   + \alpha^{zx}_{2a_{12}}z_2x_{a_12}+ \alpha^{xx}_{12} x_1x_2 + \alpha^{xx}_{1a_{11}}x_1x_{a_{11}} + \alpha^{xx}_{1a_{21}}x_1x_{a_{21}}  + \alpha^{xx}_{2a_{22}}x_2x_{a_{22}} + \alpha^{yz}_{3a_{13}}y_3x_{a_{13}} + \alpha_{4a_{23}}^{yx}y_4x_{a_{23} }
\end{align}
}

\refsec

\begin{itemize}
\item Original paper on arXiv: \cite{Kempe2004a}
\item Journal publication two years later: \cite{Kempe2006}
\end{itemize}

\newpage

\subsection{P$(3\rightarrow2)$-OT (Oliveira-Terhal, 2005)}

\summarysec

 For any 3-local term which is a product of 1-local  matrices $s_i$, we can define one auxiliary qubit labeled by $a$  and  make the transformation:


{\footnotesize
\begin{align}
a\prod_i^3 s_{i} \rightarrow \alpha   + \alpha_1^{s} s_1^2 + \alpha_2^s s_2^2 + \alpha_3^s s_3 + \alpha_a^z z_a + \alpha_{12}^{ss} s_1s_2 + \alpha_{13}^{ss} s_1^2s_3 + \alpha_{23}^{ss} s_2^2s_3 + \alpha_{3a}^{sz} s_3z_a + \alpha_{1a}^{sx}s_1x_a+\alpha_{2a}^{sx}s_2x_a   
\end{align}
}

\begin{align}
\begin{pmatrix}\alpha & \alpha_{12}^{ss}\\
\alpha_{1}^{s} & \alpha_{13}^{ss}\\
\alpha_{2}^{s} & \alpha_{23}^{ss}\\
\alpha_{3}^{s} & \alpha_{3a}^{sz}\\
\alpha_{a}^{z} & \alpha_{2a}^{sx}\\
 \textrm{N/A} & \alpha_{2a}^{sx}
\end{pmatrix}&=\begin{pmatrix}\frac{\Delta}{2} & -\Delta^{1/3}\\
-\frac{\alpha^{2/3}\Delta^{1/3}}{2} & a\frac{1}{2}\\
\frac{\alpha^{2/3}\Delta^{1/3}}{2} & a\frac{1}{2}\\
-\frac{\alpha^{1/3}\Delta^{2/3}}{2} & \frac{\alpha^{1/3}\Delta^{2/3}}{2}\\
-\frac{\Delta}{2} & -\frac{\alpha^{1/3}\Delta^{2/3}}{\sqrt{2}}\\
\textrm{N/A} & \frac{\alpha^{1/3}\Delta^{2/3}}{\sqrt{2}}
\end{pmatrix}.
\end{align}

\noindent A $k$-local Hamiltonian with a 3-local term replaced by this 2-local Hamiltonian will have an equivalent low-lying spectrum to within $\epsilon$ as long as $\Delta=\Omega\left(\epsilon^{-3}\right)$. 

\costsec

\begin{itemize}
\item 1 auxiliary qubit for each 3-local term. 
\item $\Delta =\Omega\left(\epsilon^{-3}\right)$
\end{itemize}

\examplesec

\refsec
\begin{itemize}
\item Original paper where $\alpha=1$: \cite{Oliveira2008}. For arbitrary $\alpha$ see the 2005 v1 from arXiv, or \cite{Bravyi2008}. Connection to improved version: \cite{Cao2015}.
\end{itemize}

%
\newpage

\section{Perturbative 1-by-1 Gadgets}

A 1B1 gadget allows $k$-local terms to be quadratized one step at a time, where at each step the term's order is reduced by at most one. In each step, a $k$-local term is reduced to $\left(k-1\right)$-local, contrary to SD (sub-division) gadgets which can reduce $k$-local terms to $\left(\nicefrac{1}{2}\right)$-local in one step.

\subsection{P1B1-OT (Oliveira \& Terhal, 2008)}

\summarysec

We wish to reduce the $k$-local term:

\begin{align}
H_{k-\rm{local}} = \alpha \prod_j^k s_{j} .  \label{eq:klocalInKempeMethod}
\end{align}

\noindent 

Define one auxiliary qubit labeled by $a$ and make the transformation:

\begin{align}
H_{k-{\rm local}} \rightarrow &-\left(\frac{\alpha}{2}\right)^{\nicefrac{1}{3}}\Delta^{2(1-r)}s_{k}\left(\frac{1-z_{a}}{2}\right)+\left(\frac{\alpha}{2}\right)^{\nicefrac{1}{3}}\frac{\Delta^{r}}{\sqrt{2}}\left(s_{k-1}-s_{k-2}\right)x_{a} \\
                              &+\frac{1}{2}\left(\frac{\alpha}{2}\right)^{\nicefrac{2}{3}}\left(\Delta^{r-1}s_{k-1}+{\rm sgn}(\alpha)\sqrt{2}\Delta^{-\nicefrac{1}{4}}\prod_{j}^{k-2}s_{j}\right)^{2} +\frac{\alpha}{4}\left(1+2{\rm sgn^{2}\alpha}\Delta^{\nicefrac{3}{2}-2r}\right)s_{k}.
\end{align}

\noindent The result will be a $(k-1)$-local Hamiltonian with the same low-lying spectrum as $H_{k-\rm{local}}$ to within $\epsilon$ as long as $\Delta=\Omega\left(\epsilon^{-3}\right)$.

\costsec
\begin{itemize}
\item Only 1 auxiliary qubit.
\item $\Delta =\Omega\left(\epsilon^{-3}\right)$
\end{itemize}

\examplesec

\refsec
\begin{itemize}
\item Described in: \cite{Cao2015}, based on: \cite{Oliveira2008}.
\end{itemize}

\newpage
\subsection{P1B1-CBBK (Cao, Babbush, Biamonte, Kais, 2015)}

\summarysec

Define one auxiliary qubit labeled by $a$ and make the transformation:

\begin{align}
H_{k-\rm{local}} \rightarrow &	\left(\Delta+\left(\frac{\alpha}{2}\right)^{\nicefrac{3}{2}}\Delta^{\nicefrac{1}{2}}s_{k}\right)\left(\frac{1-z_{a}}{2}\right) \\
& - \frac{\alpha^{\nicefrac{2}{3}}}{2}\left(1+{\rm sgn^{2}\alpha}\right)\left(\left(2\alpha\right)^{\nicefrac{2}{3}}{\rm sgn}^{2}\alpha+\alpha^{\nicefrac{1}{3}}s_{k}-\sqrt[3]{2}\Delta^{\nicefrac{1}{2}}\right)\left(\frac{1+z_{a}}{2}\right) \\
&	+\left(\frac{\alpha}{2}\right)^{\nicefrac{1}{3}}\Delta^{\nicefrac{3}{4}}\left(\prod_j^{k-2} s_{j}-{\rm sgn}(\alpha) s_{k-1}\right)x_{a}+{\rm sgn}(\alpha)\sqrt[3]{2}\alpha^{\nicefrac{2}{3}}\left(\Delta^{\nicefrac{1}{2}}+\Delta^{\nicefrac{3}{2}}\right)\prod_j^{k-1}s_j.
\end{align}

\noindent The result is $(k-1)$-local and its low-lying spectrum is the same as that of  $H_{k-\rm{local}}$ when $\Delta$ is large enough. 

\costsec
\begin{itemize}
\item Only 1 auxiliary qubit.
\item $\Delta =\Omega\left(\epsilon^{-3}\right)$
\end{itemize}

\examplesec

\refsec
\begin{itemize}
\item Described in: \cite{Cao2015}, based on: \cite{Oliveira2008}.
\end{itemize}
\newpage

\section{Perturbative Subdivision Gadgets}


Instead of recursively reducing $k$-local to $(k-1)$-local one reduction at a time, we can reduce $k$-local terms to $(k/2)$-local terms directly for even $k$, or to $(k+1)/2$-local terms directly for odd $k$. Since when $k$ is odd we can add an identity operator to the $k$-local term to make it even, we will assume in the following that $k$ is even, in order to avoid having to write floor and ceiling functions.

\subsection{PSD-OT (Oliveira \& Terhal, 2008)}

\summarysec

We factor a $k$-local term into a product of three factors: operators $H_1,H_2$ acting on non-overlapping spaces, and scalar $\alpha$. Then introduce an auxiliary qubit labelled by $a$ and make the transformation:

\begin{align}
H_{k-\rm{local}} \rightarrow \Delta \frac{1-z_a}{2} +  \frac{\alpha}{2}H_1^2   + \frac{\alpha}{2}H_2^2  +    \sqrt{\frac{\alpha\Delta}{2}}\left(  - H_1 + H_2   \right)x_a.
\end{align}

The resulting Hamiltonian has a degree of 1 larger than the degree of whichever factor  $H_1$ or $H_2$ has a larger degree, and the low-lying spectrum is equivalent to the original one to within $\mathcal{O}\left(\alpha\epsilon\right)$ for sufficiently large $\Delta$.

\costsec
\begin{itemize}
\item 1 auxiliary qubit for each $k$-local term that can be factored into two non-overlapping subspaces, is enough to reduce the degree down to  $\nicefrac{k}{2}+1$.
\item $\Delta =\frac{\alpha\left(||H_{(\rm{else})} + \Omega(\sqrt{2}) \textrm{max}\left( ||H_1||,||H_2||  \right) ||\right)^6}{\epsilon^2} =\Omega\left(\alpha\epsilon^{-2}\right)$.
\end{itemize}

\prossec
\begin{itemize}
\item Potentially very few auxiliary qubits needed.
\end{itemize}

\conssec
\begin{itemize}
\item Requires the ability to factor $k$-local terms into non-overlapping subspaces that are at most $\left(k-2\right)$-local in order to reduce $k$-locality. This is not possible for $z_1x_2x_3 + z_2z_3x_4$, for example. 
\item $\Delta$ needs to be rather large.
\item Cannot reduce 3-local to 2-local unless we generalize to a factor of 3 non-overlapping subspaces instead of 2. Needs to be combined with $3-\rightarrow2$ gadgets, for example.
\item A lot of work may be needed to find the optimal reduction, since each $k$-local term can be factored in many ways, and some of these ways may affect the ability to reduce other $k$-local terms.
\end{itemize}  
               
\examplesec

\refsec
\begin{itemize}
\item Original paper where $\alpha=1$: \cite{Oliveira2008}. For arbitrary $\alpha$ see the 2005 v1 from arXiv, or \cite{Bravyi2008}. Connection to improved version: \cite{Cao2015}.
\end{itemize}

\newpage

\subsection{PSD-CBBK (Cao, Babbush, Biamonte, Kais 2015)}

\summarysec

For any $k$-local term, we can subdivide it into a product of two $\left(\nicefrac{k}{2}\right)-$local terms:

\begin{align}
H_{k-\rm{local}} = \alpha H_{1,(\nicefrac{k}{2})-\rm{local}}H_{2,(\nicefrac{k}{2})-\rm{local}} + H_{(k-1)-\rm{local}}.  \label{eq:olivieraTehral}
\end{align}

\noindent Define one qubit $a$ and make the following Hamiltonian is $( \nicefrac{k}{2} )$-local:

\begin{align}
\Delta \frac{1-z_a}{2} + |\alpha|\frac{1+z_a}{2} + \sqrt{|\alpha|\Delta /2}\left({\rm sgn}(\alpha) H_{1,(\nicefrac{k}	{2}-\rm{local})} - H_{2,(\nicefrac{k}{2}-\rm{local})}\right)x_a
\end{align}

\noindent The result is a $( \nicefrac{k}{2})$-local Hamiltonian with the same low-lying spectrum as $H_{k-\rm{local}}$ for large enough $\Delta$. The disadvantage is that $\Delta$ has to be larger.

\costsec
\begin{itemize}
\item $\Delta \ge \left( \frac{2|\alpha|}{\epsilon}+1\right)(|\alpha|+\epsilon+22||H_{(k-1)-\rm{local}}  )$
\end{itemize}

\prossec
\begin{itemize}
\item only one qubit to reduce $k$ to $\lceil k/2\rceil +1$
\end{itemize}

\conssec
\begin{itemize}
\item Only beneficial for $k\ge5$.
\end{itemize}  
               
\examplesec

\refsec
\begin{itemize}
\item Original paper: \cite{Cao2015}. 
\end{itemize}

\newpage

\subsection{PSD-CN (Cao \& Nagaj, 2014)}

\summarysec

For a sum of terms that are $k$-local, with each term $j$ written as a product $H_{1j}H_{2j}$, introduce $N_{\rm core}$ `core' auxiliary qubits labeled by $a_i$ and $N_{\rm direct}$ `direct' auxiliary qubits labeled by $a_{ij}$ for each term $j$. Make all core auxiliary qubits couple to all others, and make the direct auxiliary qubits couple each $H_{1j}$ and $H_{2j}$ to the core auxiliary qubits.


\begin{equation}
\begin{array}{ccl}
\displaystyle
\sum_{j}a_j H_{1j} H_{2j} & \rightarrow & \alpha \sum_{ij} \left(1-z_{a_{ij}} z_{a_i} + \alpha_{j}x_{a_{ij}}  \left(H_{1j}  - H_{2j}\right)  \right) + \alpha \sum_i\left(1-z_{a_i} + \sum_{j}\left(1-z_{a_i} z_{a_j}\right)\right)  \\
\end{array}
\end{equation}

If we would like the spectrum of the RHS to be close to that of the LHS, with a difference of $O(\epsilon)$, then for any $d\in(0,1)$ we can choose 
\begin{align}
\label{eq:RC}
	N_{\rm{direct}} & \in \Omega\left( \max\left\{
			\epsilon^{-\frac{2}{d}}, 
			\left(\frac{\|H_\text{else}\|^2}{2M^4 \max_j|a_j|}\right)^{\frac{1}{d}},
			\left(M^3 \epsilon^{-2} \right)^{\frac{1}{1-d}}
		\right\}\right),\\ 
        \label{eq:CC}
        N_{\rm{core}} & \in \Omega \left( M^3 N_{\rm{direct}}^d \,\epsilon^{-1}\right), \\
        \alpha_{j}, \alpha & \in O(\epsilon) \label{eq:eps}
\end{align}.

\prossec
\begin{itemize}
\item For spectral error $\epsilon$, uses only $O(\epsilon)$ coupling between the qubits (See Equation \ref{eq:eps}).
\end{itemize}

\conssec
\begin{itemize}
\item Uses poly$(\epsilon^{-1})$ ancilla qubits (See Equations \ref{eq:RC} and \ref{eq:CC}).
\item The construction only describes the asymptotic scaling of the parameters rather than concrete assignments of them. 
More work is needed for finding tight non-asymptotic error bounds in perturbative expansion.
\end{itemize}

\examplesec

\refsec
\begin{itemize}
\item Original paper: \cite{Cao2015a}. 
\end{itemize}

\newpage

\newpage
\section{Perturbative Direct Gadgets}

Here we do not reduce $k$ by one order at a time (1B1 reduction) or by $\nicefrac{k}{2}$ at a time (SD reduction), but we directly reduce $k$-local terms to 2-local terms.

\subsection{PD-JF (Jordan \& Farhi, 2008)}

\summarysec 

Express a sum of $k$-local terms as a sum of products of Pauli matrices $s_{ij}$, and define $k$ auxiliary qubits laelled by $a_{ij}$ for each term $i$, and make the transformmation:
 
\begin{equation}
\sum_i \alpha_i \prod_{j}^k s_{ij} \rightarrow \frac{-k(-\epsilon)^k}{(k-1)!} \sum_i \frac{1}{2}\left( k^2 - \sum_{jl}^k  z_{a_{ij}} z_{a_{il}} \right) + \epsilon \left( \alpha_i s_{i1}x_{i1} +  \sum_{j}^k s_{ij}x_{ij} \right) -f(\epsilon)\Pi,
\end{equation}


\noindent for some polynomial $f(\lambda)$. The result is a 2-local Hamiltonian with the same low-lying spectrum to within $\epsilon^{k+1}$ for sufficiently smmall $\epsilon$.  

\costsec
\begin{itemize}
\item Number of auxiliary qubits is $tk$ for $t$ terms. 
\item Unknown requirement for $\epsilon$.
\end{itemize}

\prossec
\begin{itemize}
\item All done in one step, so easier to implement than 1B1 and SD gadgets.
\end{itemize}

\conssec
\begin{itemize}
\item Requires 2 more auxiliary qubits per term than 1B1-KKR.
\item Unknown polynomial $f(\lambda)$
\end{itemize}

\examplesec

\refsec
\begin{itemize}
\item Original paper \cite{Jordan2008}.
\end{itemize}

\newpage

\subsection{PD-BFBD (Brell, Flammia, Bartlett, Doherty, 2011)}

\summarysec 

The 4-body Hamiltonian:

\begin{eqnarray}
H_{\textrm{4-local}} = -\sum_{ij}\left( z_{4i+1,j}z_{4i+2,j}z_{4i+3,j}z_{4i+4,j} + x_{4i+3,j}x_{4i+4,j}x_{4i+6,j}x_{4i+4,j+1}  \right) 
\end{eqnarray}

\noindent is transformed into the 2-body Hamiltonian:

\begin{align}
H_{\textrm{2-local}} &= -\sum_{ij} \left( x_{8i+4,j}x_{8i+6,j} + x_{8i+3,j+1}x_{8i+5,j+1} + z_{8i+4,j}z_{8i+3,j+1} + z_{8i+6,j}z_{8i+5,j+1} +  \right.\\
&=   - \lambda\left( x_{8i+1,j}x_{8i+3,j} + x_{8i+2,j}x_{8i+4,j} + x_{8i+5,j}x_{8i+6} + x_{8i+7}x_{8i+8}\right. \\
&= \left.\left.      z_{8i+1,j}z_{8i+3,j} + z_{8i+2,j}z_{8i+4,j} + z_{8i+5,j}z_{8i+6} + z_{8i+7}z_{8i+8} \right)\right) \
\end{align} 


\noindent For $\lambda=\mathcal{O}\left(\epsilon^{-5}\right)$, the 2-local Hamiltonian has the same low-lying spectrum as the 4-local Hamiltonian, to within an error of $\epsilon$.

\costsec
\begin{itemize}
\item In total, uses four times the number of qubits of the original Hamiltonian.
\item Unknown requirement for $\lambda$.
\end{itemize}

\prossec
\begin{itemize}
\item All done in one step, so easier to implement than implementing two 1B1 gadgets.
\item Very symmetric
\end{itemize}

\conssec
\begin{itemize}
\item Ordinary 1B1 or SD followed by 3$\rightarrow$2 gadgets would require half as many total qubits.
\item Many 2-local terms.
\item Perturbative, as opposed to NR-OY which is similar but does not involve any $\lambda$ parameter.
\item Required value of $\lambda$ for it to work, is presently unknown.
\end{itemize}


\refsec
\begin{itemize}
\item Original paper: \cite{Brell2011}.
\end{itemize}

\newpage

\newpage
\part{{\normalsize {\underline{Appendix}}}}


\section{Transformations from ternary to binary variables}

\begin{eqnarray}
H_{\textrm{ternary}} = -\lambda (z_1\,z_2+z_1-z_2)   
\end{eqnarray}

In this implementation the variable $\frac{1}{2}(z_1+z_2)$ plays the role of $t$, assuming $\lambda$ is large and positive. For instance, when coupled to a binary variable $t\, z_3\rightarrow \frac{1}{2}(z_1+z_2)\,z_3$.

\newpage
\section{Further Examples}

\examplesec \label{subsec:Example_Ramsey_deduc_reduc}
Here we show how deductions can arise naturally from the Ramsey number problem.
Consider $\mathbb{\mathcal{R}}(4,3)$ with $N=4$ nodes.
Consider a Hamiltonian:
\begin{equation}
H = (1-z_{12})(1-z_{13})(1-z_{23})+\ldots+(1-z_{23})(1-z_{24})(1-z_{34})+ z_{12}z_{13}z_{14}z_{23}z_{24}z_{34}.
\end{equation}

See \cite{Okada2015} for full details of how we arrive at this Hamiltonian.

Since we are assuming we have no 3-independent sets, we know that
$(1-z_{12})(1-z_{13})(1-z_{23})=0$, so $z_{12}z_{13}z_{23}=z_{12}z_{13}+z_{12}z_{23}+z_{13}z_{23}-z_{12}-z_{13}-z_{23}+1$.
This will be our deduction.

Using deduc-reduc we can substitute this into our 6-local term to get:
\begin{eqnarray}
H & = & 2(1-z_{12})(1-z_{13})(1-z_{23})+\ldots+(1-z_{23})(1-z_{24})(1-z_{34})+\\
 &  & z_{14}z_{24}z_{34}(z_{12}z_{13}+z_{12}z_{23}+z_{13}z_{23}-z_{12}-z_{13}-z_{23}+1).
\end{eqnarray}

We could repeat this process to remove all 5- and 4-local terms without adding any auxiliary qubits.
Note in this case the error terms added by deduc-reduc already appear in our Hamiltonian.




\newpage
\section{$2\rightarrow 2$ gadgets}
This review has only focused on $k-$local to $2-$local transformations where $k>2$. There is also a large number of $2-$local to $2-$local transformations in the literature, which are used for various pruposes. Some of these are listed here:

\vspace{10mm}

\begin{itemize}
\item Gadgetization of any $2-$local Hamiltonian into $\{\openone,z,x,zz,xx\}$ or $\{\openone,z,x,zx\}$ \cite{Biamonte2008}. Used for the proof that $xx + zz$ or $xz$ is universal is enough for universal quantum computation. In other words, \textit{any} computation can be transformed into a problem of finding the ground state of a $2-$local Hamiltonian containing terms from $\{\openone,z,x,zz,xx\}$ or from $\{\openone,z,x,zx\}$ with real coefficients, and the ground state can be found by adiabatic quantum computing with only polynomial time and space overhead over the best alternative algorithm for the problem.

\item Transformation of any $2-$local Hamiltonian into $\{\openone,z,x,zz,xx+yy\}$, without any perturbative gadgets, and only requiring the qubits to be connected in an almost 2D lattice \cite{Lloyd2016}.

\item "Cross gadget", "fork gadget", and "triangle gadget" described in \cite{Oliveira2008}.
\item Gadgetizeation of a $2-$local Hamiltonian with very strong couplings, into a $2-$local Hamiltonian with strengths in $\mathcal{O}\left( \nicefrac{1}{\textrm{poly}\left( \epsilon^{-1},n\right)}  \right)$, and $\textrm{poly}\left( \epsilon^{-1},n  \right)$ auxiliary qubits and $\textrm{poly}\left( \epsilon^{-1},n  \right)$ new quadratic terms. \cite{Cao2015a}.
\item $yy$ creation gadget: Simulation of $yy$ terms  using $\{\openone,z,x,zz,xx\}$, with coupling strength restriction defined according to $\Delta = \Theta\left(\epsilon^{-4}\right)$ \cite{Cao2015}. 
\end{itemize}

\subsection{Minor-embedding quadratic functions for different graphs}
\begin{itemize}
\item Minor-embedding general problems for the Chimera  \cite{Neven2009} graph \cite{vchoi08,Choi2011a}.
\item Minor-embedding quadratization gadgets for the Pegasus  \cite{Dattani2019b} graph \cite{Dattani2019c}.
\end{itemize}

\newpage

\section{Further References}
\begin{itemize}
\item Gadgets for pseudo-Boolean optimization problems, with reduced precision requirements: \cite{Babbush2013}.
\item Formalization of pseudo-Boolean gadgets in quantum language. \cite{Biamonte2008a}.
\item By adding more couplers and more auxiliary qubits, we can bring the error down arbitrarily low:  \cite{Cao2015a}.
\item More toric code gadgets: \cite{Brell2014}.
\item Parity adiabatic quantum computing (LHZ lattice): \cite{Lechner2015}.
\item Extensions of the LHZ scheme: \cite{Rocchetto2016}.
\item Minimizing $k$-local discrete functions with the help of continuous variable calculus \cite{Shen2017a}.
\item ORI graph which attempts to give optimal quadratizations \cite{Gallagher2011}.
\item Survey on pseudo-Boolean optimization \cite{Boros2002}.
\item Linearization of equations before they are squared \cite{Schaller2010}, and its application to factoring numbers \cite{Xu2012}.
\item Mentioned in \cite{Ali2008} as an early application of quadratization: \cite{Cunningham1985}.
\item Characterizaton of NTRs for cubics: \cite{Crama2014a}.
\item Relation between cones of nonnegative quadratic pseudo-Boolean functions and the Boolean quadric polytope \cite{BorosLari2014}.
\item Relation between cones of nonnegative quadratic pseudo-Boolean functions and the Boolean quadric polytope \cite{BorosLari2014}.
\item Effective non-Hermitian Hamiltonian with 3-body interactions which helps to calcualte electronic structure energies closer to the complete basis set limit \cite{Cohen2019}.
\end{itemize}

\newpage
\section{Circuits that effectively implement degree-$k$ terms for superconducting qubits}
\begin{itemize}
\item Presentation by Northrop Grumman about a $zzz$ coupler \cite{Strand2017} and associated patent \cite{Ferguson2017,Ferguson2018}.
\item Presentaiton that included discussion about engineering multi-qubit interactions \cite{Kerman2018}, presentation by the same lab about the design and experimental demonstration of a $zzzz$ coupling \cite{Menke2019}, and associated patent \cite{Kerman2018a,Kerman2019,Kerman2019a}.
\item Design of an effective $zzzz$ coupling without any auxiliary logical qubits \cite{Schondorf2018}.
\item Design of a tunable $zzz$ coupling in which all $zz$ couplings are cancelled, and its experimental demonstration \cite{Melanson2019}.
\end{itemize}

\newpage


\section{Contributors}

\subsection*{Richard Tanburn}
\begin{itemize}
\item Richard was the original creator and maintaner of the Git repository.
\item Richard created the Tex commands used throughout the document, and contributed majorly to the overall layout.
\item Richard wrote the original versions of the following sections: (1) Deduc-Reduc, (2) ELC Reduction, (3) Groebner Bases, (4) Split Reduction, (5) NTR-KZFD, (6) NTR-GBP, (7) PTR, (8) PTR-Ishikawa, (9) PTR-KZ, (10) PTR-GBP, (11) Bit flipping, (12) RBS, and (13) FGBZ. 
\item Richard also wrote the "Further Example" of Deduc-Reduc in the Appendix.
\end{itemize}

\subsection*{Nicholas Chancellor}
\begin{itemize}
\item Nick made contributions to the following sections: (1) RMS (in terms of z), (2) PTR-RBL-(3$\rightarrow$2), (3) PTR-RBL-(4$\rightarrow$2), (4) SBM, (5) Flag based SAT Mapping, and to the qutrit $\rightarrow$ qubit transformation (6).
\end{itemize}

\subsection*{Szilard Szalay}
\begin{itemize}
\item Szilard re-derived Nike's transformations for the sections: (1) SFR-BCR-1, (2) SFR-BCR-2, (3) SFR-BCR-3, and (4) SFR-BCR-4 from the notation of the original paper, into the format consistent with the rest of the book. In doing so he corrected errors in Nike's work and also fixed them in the main document.
\end{itemize}

\subsection*{Ka Wa Yip}
\begin{itemize}
\item Ka Wa Yip added the page about his own method co-authored with Xu, Koenig and Kumar.
\end{itemize}

\subsection*{Yudong Cao}
\begin{itemize}
\item Yudong wrote the first version of the following section: (1) PSD-CN.
\end{itemize}

\subsection*{Daniel Nagaj} 
\begin{itemize}
\item Daniel provided a .tex document to Nike in May 2018 which helped Nike to write the following sections: (1) NP-Nagaj-1, (2) NP-Nagaj-2. The document that Daniel provided made it easier for Nike to write these sections than the original papers.
\end{itemize}

\subsection*{Aritanan Gruber} 
\begin{itemize}
\item Aritanan informed us in August 2015 of what we ended up making the following sections: (1) PTR-Ishikawa. 
\end{itemize}

\subsection*{Charles Herrmann} 
\begin{itemize}
\item Charles informed us in May 2018 of the papers which contained results which became the following sections: (1) PTR-BCR-1,  (2) PTR-BCR-2, (3) PTR-BCR-3, (4) PTR-BCR-4, (5) SFR-BCR-1, (6) SFR-BCR-2, (7) SFR-BCR-3, (8) SFR-BCR-4, (9) SFR-BCR-5, (10) SFR-BCR-6. 
\end{itemize}

\subsection*{Elisabeth Rodriguez-Heck} 
\begin{itemize}
\item Elisabeth provided us with a 2-page PDF document with valuable comments on the entire Book.
\item Elisabeth also pointed us to what became the following section: (1) ABCG Reduction.
\end{itemize}

\subsection*{Hou Tin Chau}
\begin{itemize}
\item Tin made the examples for the following sections: SFR-BCR-1,2,3,4.
\item Tin fixed a typo in the alternative forms of the following sections: SFR-BCR-3,4.
\end{itemize}

\subsection*{Andreas Soteriou}
\begin{itemize}
\item Andreas found typos on the opening page in the arXiv version which surprisingly no one else found (or pointed out), and he diligently fixed them.
\item Andreas created the example involving $x$,$y$, and $z$ presented on the opening page in the September 2019 version (I plan to have this example further improved at a later time). 
\end{itemize}

\subsection*{Jacob Biamonte} 
\begin{itemize}
\item Jacob made valuable edits during a proof-reading of the book.
\end{itemize}

\newpage

\section*{Acknowledgments} 
\begin{itemize} 
\item It is with immense pleasure that we thank Emile Okada of Cambridge University, who during his first year of udnergraduate study, worked with Nike Dattani and Richard Tanburn on quadratization of pseudo-boolean functions for quantum annealing, and in the first half of 2015 played a role in the development of the Deduc-Reduc and Split-Reduc and Groebner bases methods presented in this review.
\item We thank Gernot Schaller of University of Berlin, who in December 2014 provided Nike Dattani with insights into his quadratization methods mentioned in this review paper, as well as for sharnig his Mathematica code which could be used to generate such quadratization formulas and others.
\item We thank Mohammad Amin of D-Wave for pointing Nike Dattani to the paper \cite{Bian2013} on determining Ramsey numbers on the D-Wave device, which contained what we call in this review "Reduction by substitution", later found through Ishikawa's paper to be from the much older 1970s paper by Rozenberg.
\item We thank Catherine McGeoch of Amherst University and D-Wave, for helpful discussions with Nike Dattani in December 2014 about how to map quadratic pseudo-Boolean optimization problems onto the chimera graph of the D-Wave hardware and for pointing us to the important references of Vicky Choi. While chimerization is very different from quadratization, understanding that roughly $n^2$ variables would be needed to map a quadratic function of $n$ variables, helped Nike Dattani and Richard Tanburn to appreciate how impotrant it is to be able to quadratize with as few variables as possible, and having this in mind throughout our studies helped inspire us in our goals towards "optimal quadratization".
\item We thank Aritanan Gruber and Endre Boros of Rutgers University, who in August 2015 shared with Nike Dattani and Richard Tanburn some of their wisdom about sub-modularity, and Aritanan Gruber for pointing us to the then very recent paper of Hiroshi Ishikawa on what we call in this review "ELC reductions", which was also a valuable paper due to the references in it. We also thank him for helping us in our quest to determine whether or not "deduc-reduc" was a re-discovery by Richard, Emile, and Nike, or perhaps a novel quadratization scheme.
\item We thank Toby Cathcart-Burn of Oxford University, who during the third year of undergraduate study, worked with Nike Dattani and Richard Tanburn and in Autumn 2015 and Winter 2016 helped us gain insights about the application of deduc-reduc and bit flipping to the problem of determining Ramsey numbers via discrete optimization, and for insights into the trade-offs between Ishikawa's symmetric reduction and reduction by substitution. 
\item  We thank Hiroshi Ishikawa of Waseda University, who Nike Dattani had the memorable opportunity to visit in November 2015, and through discussions about the computer vision problem and neural network problem (two examples of real-world discrete optimization problems that benefit from quadratization), provided insights about the role of quadratization for calculations on classical computers. In particular, solving the computer vision problem in which he had experience solving on classical computers, was very different from the integer factoring problem and Ramsey number problem which we had been attempting to quadratize for D-Wave and NMR devices. He taught us that far more total (original plus auxiliary) variables can be tolerated on classical computers than on D-Wave machines or NMR systems, and approximate solutions to the optimization problems are acceptable (unlike for the factorization and Ramsey number problems in which we were interested). This gave us more insight into what trade-offs one might wish to prioritize when quadratizing optially. We also thank him for helping us in our quest to determine whether or not "deduc-reduc" was a re-discovery by Richard, Emile, and Nike, or perhaps a novel quadratization scheme.
\item Last but indubitably not least, we thank Jacob Biamonte of Skolkovo Institute of Technology, who Nike Dattani enjoyed visiting in Hangzhou in January 2017 and meeting at Harvard University in April 2018. Jacob provided us plenty of insights about perturbative gadgets, made valuable comments on early versions of our manuscript, and has been a major supporter of this review paper since the idea was presented to him in December 2016. At many points during the preparation of this review, we had prioritized other commitments and put preparation of this review aside. Jake's frequent encouragement was often what got us working on this review again. Without him, we are not certain this paper would have been completed by this time (or ever!).


\end{itemize}

\newpage
\bibliography{k-local-quadratization}

\end{document}